\documentclass{article}
\usepackage{graphicx} 
\usepackage{amsthm}
\usepackage{amsmath, amsfonts, amssymb, bbm, bm}
\usepackage{mathabx}
\usepackage{graphicx}
\usepackage{appendix}
\usepackage[utf8]{inputenc}
\usepackage[T1]{fontenc}
\usepackage[numbers, sort&compress]{natbib}
\usepackage{enumerate}
\usepackage{xcolor}
\usepackage{url}
\usepackage{float}
\usepackage{placeins}
\usepackage[a4paper, left=2cm, right=2cm, top=2cm, bottom=2cm]{geometry}

\theoremstyle{plain}
\newtheorem{defi}{Definition}
\newtheorem{theorem}{Theorem}
\newtheorem{ex}{Example}
\newtheorem{lem}{Lemma}

\newtheorem{ass}{Assumption}
\newtheorem{rem}{Remark}

\newtheorem{cor}{Corollary}

\newcommand{\R}{\mathbb{R}}
\newcommand{\T}{\mathcal{T}}
\newcommand{\E}{\mathrm{E}}
\renewcommand{\P}{P}
\newcommand{\Var}{\mathbb{V}ar}

\title{General Methodology for Asymptotically Balanced Multiple Tests}
\author{Merle Munko$^{1}$}
\date{}

\begin{document}

\maketitle
\begin{center}\noindent${}^{1}$ {NOVA Information Management School (NOVA IMS), Universidade NOVA de Lisboa, Campus de Campolide, 1070-312 Lisboa, Portugal}

\vspace*{8.25ex}
 \end{center}

\noindent \textbf{Abstract:}
In statistical applications, often more than one hypothesis is of interest and, thus, multiple testing problems occur naturally. To control the family-wise error rate, popular adjustment methods, like the Bonferroni and Holm correction, have been developed. However, these methods do not use the dependence structure of the test statistics. Consequently, they are conservative in most settings. To increase the power of the tests, the joint dependence structure of the test statistics can be incorporated for the construction of critical values, for example, by using resampling schemes as bootstrap procedures to approximate the joint distribution. This has already been applied in several contexts. In this work, we develop a very flexible and general theoretical foundation, which can be used to construct asymptotically balanced multiple tests that control the family-wise error rate in the strong sense. Additionally, we prove in the general setting that consistent resampling schemes satisfy the needed conditions.


\section{Introduction}\label{sec:Intro}
Statistical hypothesis tests offer a technique to make decisions about a null hypothesis.
Tests are required to control the type-I error rate, that is, the probability that the null hypothesis is rejected although the null hypothesis is true.
However, often more than one hypothesis is of interest in practice and, hence, more than one test is needed to infer the hypotheses simultaneously.
By just controlling the type-I error rate for each test, the probability that at least one true null hypothesis is rejected generally increases.
In contrast, we aim to control the family-wise error rate (FWER), i.e., the probability of rejecting at least one true null hypothesis, in multiple testing problems.

There are some well-known p-value adjustments like the Bonferroni correction and the Holm correction that ensure FWER control in the strong sense, i.e., under any combination of true and false hypotheses, and are applicable for all underlying dependence structures of the tests \cite{dunn_multiple_1961,Holm}.
Furthermore, the Bonferroni correction naturally guarantees a balancedness condition:
The local type-I error of each test is 
controlled by the same value, say $\alpha/L$, where $\alpha$ is the global level of significance and $L$ is the number of tests.

However, when more information about the test's dependence is known, more powerful test decisions can be achieved than those obtained from the Bonferroni or Holm correction. 
One approach is to perform so-called max-type tests that use the 
distribution of the maximum of all test statistics for deriving critical values \cite{hothorn,Romano01032005,westfall2008multiple,westfall1993resampling}.
While this methodology exploits the global level of significance, 
it generally does not retain the balancedness condition if the marginal distributions differ, i.e., if the test statistics have different distributions.

In this paper, we derive a general methodology for deriving powerful, asymptotically balanced multiple tests.
The main theorem of this work (Theorem~\ref{BSQuant}) states that the constructed multiple tests control the FWER in the strong sense and are asymptotically balanced.
The ideas in this paper have already been applied in rather specific scenarios for multiple testing problems, see, e.g., \cite{munko2023rmstbased, munko2025FMANOVA, munkoRMTL, Munko2025AStA, sattler2025quadraticformbasedmultiple}. However, in the present work, we formulate the approach in a general way such that asymptotically balanced multiple tests can easily be applied to other multiple testing problems.
We want to emphasize the generality of the proposed methodology by highlighting that it is not tailored to specific hypotheses, testing problems, test statistics, or models.
Additionally, we explicitly allow non-measurable test statistics, thereby broadening applicability further.

In the literature \cite{beran1988}, (asymptotically) balanced simultaneous confidence sets were already constructed in a general setting. However, Assumptions~1 and~2 in \cite{beran1988} require convergence results over probability measures $P_{\theta_n,n}$ for all converging sequences $\theta_n$ of parameters. This condition seems to be hard to verify in various applications, especially in non-parametric problems, since many classical limit theorems require a fixed underlying probability measure.
In contrast, our main theorem (Theorem~\ref{BSQuant}) requires weakly consistent approximations of the joint and marginal distribution functions of the test statistics, cf. \eqref{eq:condition} and \eqref{eq:condition2}. 
As such, the conditions are easily understandable and interpretable, and often result naturally from classical limit theorems. 

Resampling schemes like nonparametric, parametric, and wild bootstrap approaches are popular tools for approximating the distributions of test statistics.
Therefore, we verify that the conditions \eqref{eq:condition} and \eqref{eq:condition2} hold for consistent resampling schemes and for the corresponding Monte Carlo approximations of the distribution functions (Lemmas~\ref{ResamplingLemma} and~\ref{MonteCarloRemark}). 

The remainder of this paper is organized as follows:
Firstly, we present the multiple testing setup in Section~\ref{sec:MultiSetup}.
The choice of suitable critical values is studied in Section~\ref{sec:CritVal}.
The main theorem in Section~\ref{sec:Main} provides a multiple testing procedure that controls the FWER in the strong sense asymptotically.
Moreover, we show that the critical values for the tests can be determined by consistent resampling schemes in Section~\ref{sec:Consistent}.
Finally, Section~\ref{sec:Discussion} summarizes and discusses the results of this work.

\section{General multiple testing setup}\label{sec:MultiSetup}
In this paper, we consider a general multiple testing problem with $L\in\mathbb N$ local null hypotheses $$\mathcal{H}_{0,1},\quad \ldots, \quad  \mathcal{H}_{0,L}$$ associated with sequences of test statistics $W_{1,n},\ldots,W_{L,n}$. Throughout this work, $n\in\mathbb N$ usually represents the sample size, and we consider convergence results as $n\to\infty$.
We assume that {large} values of the $\ell$th test statistic $W_{\ell,n}$ indicate a rejection of the $\ell$th null hypothesis $\mathcal{H}_{0,\ell}$, for each $\ell\in\{1,\dots,L\}$. In Section~\ref{sec:CritVal}, we will discuss how \textit{large} can be quantified, i.e., how the critical values for the local test decisions can be constructed. 

However, we first need to introduce some further notations and explanations.
Let $(\Omega,\mathcal A, P)$ be the underlying probability space.
Here, $P$ is an unknown probability measure from a class $\mathcal P$ of probability measures on $(\Omega, \mathcal A)$.
No further assumptions on $\mathcal P$ are required at this point. Thus, we explicitly include non-parametric as well as parametric models.

The $\ell$th hypothesis states that the underlying probability measure $P$ lies in a subset $\mathcal P_\ell \subset \mathcal P $, i.e., $\mathcal H_{0,\ell}: P \in \mathcal P_\ell$ for all  $\ell\in\{1,\ldots,L\}$.
In the following, if we say that a statement holds ``under $\bigcap_{\ell\in\T}\mathcal{H}_{0,\ell}$'' for some $\T \subset \{1,\ldots,L\}$, we mean that it holds for $P \in \bigcap_{\ell\in\T}\mathcal{P}_{\ell}$.
The global null hypothesis is the intersection of all local hypotheses, that is
$$ \mathcal{H}_0 = \bigcap_{\ell=1}^L \mathcal H_{0,\ell}: P \in \bigcap_{\ell=1}^L\mathcal P_\ell .$$

Additionally, despite the common definition of test statistics, we do not assume measurability of the maps $$W_{1,n},\ldots,W_{L,n}: \Omega \to \R$$ in the following.
Although in most applications the measurability of the test statistics will be given, non-measurable test statistics might be of interest, for example a Kolmogorov-Smirnov statistic of empirical processes over an uncountable function class; see Section~3.8 in \cite{vaartWellner2023}.


The following example serves to illustrate the previously introduced quantities.

\begin{ex}
Let us assume that we have $k \in\mathbb N\setminus\{1\}$ independent samples of independent and identically distributed (i.i.d.) random variables $X_{i1},\ldots,X_{in_i} \sim P_i, i\in\{1,\ldots,k\}$, where $P_1,\ldots,P_k$ are unknown distributions on $(\R, \mathcal{B}(\R))$ with existing second moments, i.e., $\E(X_{i1}^2) = \int_{\R} x^2 \;\mathrm{d}P_i(x) < \infty$. Here, $\mathcal{B}(\R)$ denotes the Borel $\sigma$-algebra on $\mathbb R$.
Moreover, we denote the means by $\mu_i := \E(X_{i1})$ 
and the variances by $\sigma_i^2 := \Var(X_{i1}) 
, i\in\{1,\ldots,k\}$. Additionally, we assume $\sigma_1^2 > 0$.

In this example, the sample space $\Omega = \left(\R^{\mathbb N}\right)^k$ is the Cartesian product of the real numbers equipped with the product $\sigma$-algebra $\mathcal A =  \left((\mathcal{B}(\R))^{\otimes \mathbb N}\right)^{\otimes k}$.
    As class of probability measures, we can consider the class 
  \begin{align*}
      \mathcal{P} := \left\{\bigotimes_{i=1}^k P_i^{\otimes \mathbb N} \mid P_i \text{ is a probability measures on } (\R,\mathcal{B}(\R)) \text{ with } \int_{\R} x^2 \;\mathrm{d}P_i(x) < \infty \text{ for all }i\in\{1,\ldots,k\} \right. \\ \left. \text{ and } \int_{\mathbb R} \left( y - \int_{\mathbb{R}} x\,\mathrm{d}P_1 \right)^2 \,\mathrm{d}P_1(y) > 0\right\}
  \end{align*}    
    of product probability measures on $(\Omega, \mathcal A)$ with existing second moments.
    By defining the subsets
    $$ \mathcal{P}_\ell := \left\{ \bigotimes_{i=1}^k P_i^{\otimes \mathbb N} \in\mathcal P \mid  \int_{\R} x \;\mathrm{d}P_{\ell + 1}(x) = \int_{\R} x \;\mathrm{d}P_{1}(x) \right\} \subset \mathcal P, \quad \ell\in\{1,\ldots,k-1\}, $$
    we obtain the $L =k-1$ local null hypotheses
    $$\mathcal{H}_{0,1}:\mu_2 = \mu_1,\quad \ldots, \quad  \mathcal{H}_{0,k-1}: \mu_k = \mu_1.$$
    Hence, the $\ell$th local null hypothesis describes the equality of the $(\ell+1)$th mean $\mu_{\ell + 1}$ and the mean $\mu_1$.
   As the intersection of all local hypotheses, the global null hypothesis $\mathcal{H}_0: \mu_1 = \ldots = \mu_k$ describes the equality of all means.

Different test statistics can be used for testing $\mathcal H_{0,\ell}: \mu_{\ell + 1} = \mu_1$. 
One option is to use the Welch t-test statistics
    \begin{align}
        \label{eq:stud}
        W_{\ell,n} = \frac{|\bar{X}_{\ell + 1} - \bar{X}_{1}|}{\sqrt{S^2_{\ell+1}/n_{\ell + 1} + S^2_{1}/n_1}}, \quad \ell\in\{1,\ldots,k-1\},
    \end{align}
   with $n := \sum_{i=1}^k {n_i}$, $\bar{X}_{i} := \frac{1}{n_i}\sum_{j=1}^{n_i} X_{ij}$, and  $S^2_i := \frac{1}{n_i - 1} \sum_{j=1}^{n_i}(X_{ij} - \bar{X}_i)^2$ for all $i\in\{1,\ldots,k\}$.
    Another option is to use the unstudentized test statistics
    \begin{align}\label{eq:unstud}
        W_{\ell,n} = \sqrt{n}{|\bar{X}_{\ell + 1} - \bar{X}_{1}|}, \quad \ell\in\{1,\ldots,k-1\}.
    \end{align}
\end{ex}

In order to use the joint distribution of the test statistics,
we will access it through asymptotic theory. Therefore, let $\mathrm{CDF}(\R^L), \mathrm{CDF}(\R)$ denote the set of all cumulative distribution functions on $\R^L,\R$, respectively. We assume the following convergence result for the joint distribution of the test statistics.

\begin{ass}[Joint convergence of the test statistics]\label{Wass}
Let $(W_1,\ldots,W_L)$ denote a random vector with cumulative distribution function $\mathcal F \in\mathrm{CDF}(\R^L)$ and continuous marginal cumulative distribution functions $\mathcal F_1,\ldots,\mathcal F_L\in\mathrm{CDF}(\R)$, i.e., $\mathcal F_\ell$ denotes the cumulative distribution function of $W_\ell$ for each $\ell \in \{1,\ldots,L\}$.
 Moreover, let $(W_{1,n},\ldots,W_{L,n}), {n\in\mathbb N},$ be a sequence of $L$ test statistics
    and suppose that
\begin{align}\label{eq:DistrConv}
    \left(W_{\ell,n}\right)_{\ell\in\T}  \xrightarrow{d} \left(W_\ell\right)_{\ell\in\T} \quad\text{as  $n\to\infty$   under $\bigcap_{\ell\in\T} \mathcal{H}_{0,\ell}$ for all index sets $\T\subset \{1,\dots,L\}$,}
\end{align}
where convergence in distribution is understood in the sense of \cite{vaartWellner2023}.
\end{ass}

The assumption states that the vector of test statistics converges in distribution to a random vector $(W_1,\ldots,W_L)$ with continuous marginal distributions under the global null hypothesis.
In addition, if only some of the local null hypotheses hold, say $\mathcal H_{0,\ell}, \ell \in \mathcal T \subset \{1,\ldots,L\}$, we still require distributional convergence of the vector of corresponding test statistics $(W_{\ell,n})_{\ell\in\mathcal T}$ to the marginal $(W_\ell)_{\ell\in\mathcal T}$ of the limit vector above.

In Example~\ref{myEx}, the assumption can be shown to be valid.

\addtocounter{ex}{-1}
\begin{ex}[continued]\label{myEx}
In the following, we further assume that no sample size vanishes asymptotically, i.e., $n_i/n \to \kappa_i \in (0,1)$ holds for all $i\in\{1,\ldots,k\}.$
Then, let $(Z_1,\ldots,Z_{k-1})$ be a centered normally distributed random vector with covariance matrix
    \begin{align}\label{eq:covmat}
   \boldsymbol{\Sigma} := \kappa_1^{-1}\sigma_1^2 \cdot \mathbf{1}_{(k-1)\times (k-1)} + \mathrm{diag}(\kappa_2^{-1}\sigma_2^2, \ldots, \kappa_k^{-1}\sigma_k^2) 
   ,
\end{align}
where $\mathbf{1}_{(k-1)\times (k-1)} \in \mathbb{R}^{(k-1)\times (k-1)}$ denotes the matrix with all entries equal to one and $\mathrm{diag}(\kappa_2^{-1}\sigma_2^2, \ldots, \kappa_k^{-1}\sigma_k^2)$ denotes the diagonal matrix with diagonal entries $\kappa_2^{-1}\sigma_2^2, \ldots, \kappa_k^{-1}\sigma_k^2$.

    The central limit theorem provides that \eqref{eq:DistrConv} holds for the proposed test statistics:
    \begin{itemize}
        \item For the Welch t-test statistics \eqref{eq:stud}, we obtain the limit variable $$(W_1,\ldots,W_{k-1}) = \left(\frac{|Z_1|}{\sqrt{\kappa_1^{-1}\sigma_1^2 + \kappa_2^{-1}\sigma_2^2}}\ ,\ \ldots\ ,\ \frac{|Z_{k-1}|}{\sqrt{\kappa_1^{-1}\sigma_1^2 + \kappa_k^{-1}\sigma_k^2}}\right).$$
        It can be noted that the marginal distributions are folded standard normal distributions.
        \item For the unstudentized test statistics \eqref{eq:unstud}, we obtain the limit variable
    $(W_1,\ldots,W_{k-1}) = (|Z_1|,\ldots,|Z_{k-1}|)$.
    \end{itemize}
\end{ex}

Next, we aim to choose suitable critical values based on the joint (limit) distribution from Assumption~\ref{Wass} for the test statistics, which is investigated in the next section.

\section{Choice of the critical values}\label{sec:CritVal}
In this section, we discuss approaches to how the joint (limit) distribution of a vector of test statistics can be used to determine suitable critical values.
In particular, we firstly describe max-type critical values and their imbalancedness. After defining the balancedness condition, we turn to the construction of critical values that yield asymptotically balanced multiple tests.

\textbf{Imbalancedness of max-type critical values} A classical approach for compatible local and global test decisions would be to calculate a (max-type) critical value for the maximum statistic $\max_{\ell\in\{1,\ldots,L\}} W_{\ell,n}$. Then, the $\ell$th null hypothesis $\mathcal{H}_{0,\ell}$ is rejected whenever the corresponding local test statistic $W_{\ell,n}$ exceeds the critical value. 
In the special case that the distributions of $W_1,\dots,W_L$ in Assumption~\ref{Wass} are equal, as, e.g., for the Welch t-test statistics in Example~\ref{myEx}, this results asymptotically in the same probability of being wrongly rejected for each local hypothesis.

However, if the marginal distributions are not equal, the local hypotheses may have different asymptotic probabilities of being wrongly detected by using max-type critical values and, thus, are not treated in the same way. 
\begin{figure}[b]
    \centering
    \includegraphics[trim={0 2.3cm 1cm 2.1cm},
    clip, width=\textwidth]{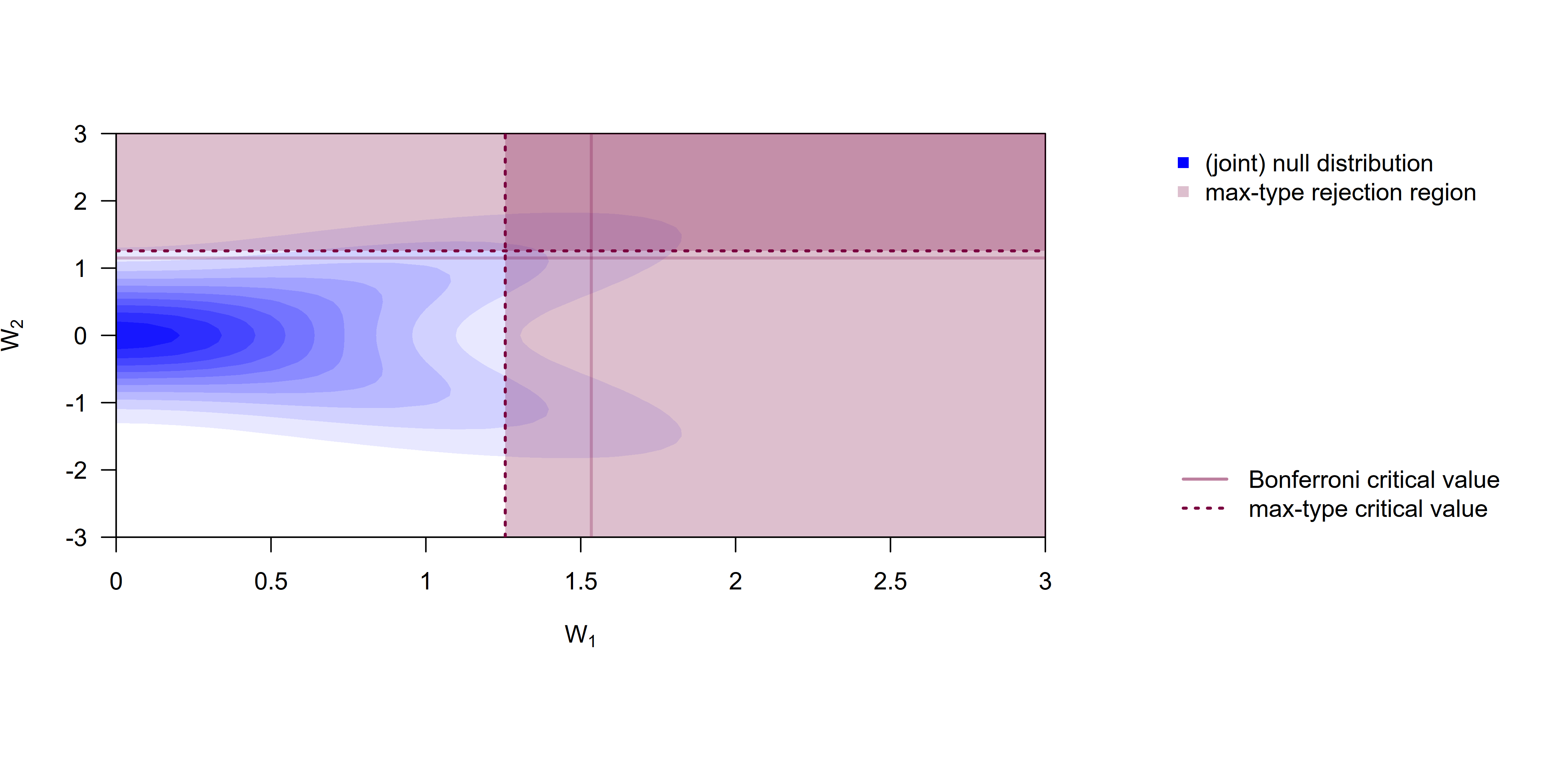}
    \caption{Exemplary illustration of the max-type testing procedure (dotted lines) for two test statistics with different distributions compared to the Bonferroni correction (solid lines).}
    \label{fig:maxtype}
\end{figure}
While in some cases, equal marginal distributions can be obtained by a simple studentization, cf. Example~\ref{myEx}, this is not always the case, as the following example illustrates.

\begin{ex}\label{myEx2}
    Let us assume that we have $3$ independent samples of i.i.d. random variables $X_{i1},\ldots,X_{in_i} \sim P_i, i\in\{1,2,3\},$ where $P_1,P_2,P_3$ are unknown distributions on $(\R, \mathcal{B}(\R))$ with existing second moments. We denote $\mu_i = \E(X_{i1})$ for all $i\in\{1,2,3\}$.
    Now, we aim to test the two hypotheses
    $$  \mathcal{H}_{0,1}: \mu_2 = \mu_1 \quad\text{and}\quad \mathcal{H}_{0,2}: \mu_3 \leq \mu_1 $$
    simultaneously, i.e., we are interested in a two-sided comparison of the means of the first two groups and a one-sided comparison of the means of group $1$ and $3$.
    Here, suitable test statistics are
    \begin{align*}
        W_{1,n} = \sqrt{n}{|\bar{X}_{2} - \bar{X}_{1}|} \quad \text{and} \quad W_{2,n} = \sqrt{n}({\bar{X}_{3} - \bar{X}_{1}})
    \end{align*} with the same notation as in Example~\ref{myEx}.
    By the central limit theorem, we obtain
    $(W_{1,n}, W_{2,n}) \xrightarrow{d} (|Z_1|, Z_2)$ as $n\to\infty$, where $(Z_1,Z_2)$ is as in Example~\ref{myEx}.

    Hence, we observe that the marginal distributions are not equal: While the first test statistic $W_{1,n}$ converges weakly to a folded normal distribution, the second test statistic $W_{2,n}$ converges weakly to a normal distribution.
    Even a standardization of the test statistics can not yield equal marginal limit distributions in this case.
\end{ex}

Figure~\ref{fig:maxtype} shows an exemplary illustration of the density of the joint distribution of $(W_1, W_2) = (|Z_1|, Z_2)$ from Example~\ref{myEx2} as well as the corresponding Bonferroni and max-type critical values.
For two hypotheses, the Bonferroni critical values are the $(1-\alpha/2)$-quantiles of the marginal distributions.
The figure illustrates that using max-type critical values instead leads to notably lower critical values for the first test statistic $W_{1,n}$ and, thus, larger rejection regions and a higher power of the first test.
However, when we compare the critical values for the second test statistic $W_{2,n}$, we observe a different effect:
The Bonferroni critical value is slightly lower than the max-type critical value. 
In particular, the resulting max-type test is slightly less powerful than the Bonferroni-corrected test for $\mathcal H_{0,2}$.
Additionally, we can conclude that the two hypotheses are not treated in the same way by the max-type tests:
While the asymptotic type I error for testing the first hypothesis clearly exceeds $\alpha/2$,
the asymptotic type I error for testing the second hypothesis is slightly less than $\alpha/2$.
This imbalancedness results from the different marginal distributions of the joint limit distribution.

\textbf{Asymptotically balanced multiple tests}
To treat unequal marginal distributions in a fair way, we adopt the idea of balanced simultaneous confidence sets as in \cite{beran1988}.
In detail, we aim to find individual critical values $q_{1,n},\ldots,q_{L,n}$ for the $L$ test statistics such that
\begin{align}\label{eq:(i)}
    \limsup_{n\to\infty} \P^*\left(\exists \ell\in\T :\: W_{\ell,n} >  q_{\ell,n}\right) \leq \alpha\quad\text{under $\bigcap_{\ell\in\T} \mathcal{H}_{0,\ell}$ for all index sets $\T\subset \{1,\dots,L\}$}
\end{align}
for a global level $\alpha\in (0,1)$ and, for some $\beta \in (0,\alpha]$,
\begin{align}\label{eq:(ii)}
    \limsup_{n\to\infty}\P^*\left(W_{\ell,n} >  q_{\ell,n}\right) = \beta\quad\text{under $\mathcal{H}_{0,\ell}$ for all $\ell\in\{1,\dots,L\}$},
\end{align}where $\beta$ does not depend on $\ell$.
Here and throughout, $\P^*$ denotes the outer probability to avoid measurability issues. By doing so, we still allow for non-measurable $W_{\ell,n}$ and $q_{\ell,n}$.

The first condition \eqref{eq:(i)} ensures the asymptotic FWER control in the strong sense.
Here, ``{in the strong sense}'' means that the FWER is controlled for any set $\T$ of true null hypotheses.
The second condition \eqref{eq:(ii)} guarantees that all local hypotheses are treated in the same way and is referred to as \textit{asymptotically balanced} multiple tests, cf. \cite{beran1988}.

\begin{figure}[bt]
    \centering
    \includegraphics[trim={0 2.3cm 1cm 2.1cm},
    clip, width=1\textwidth]{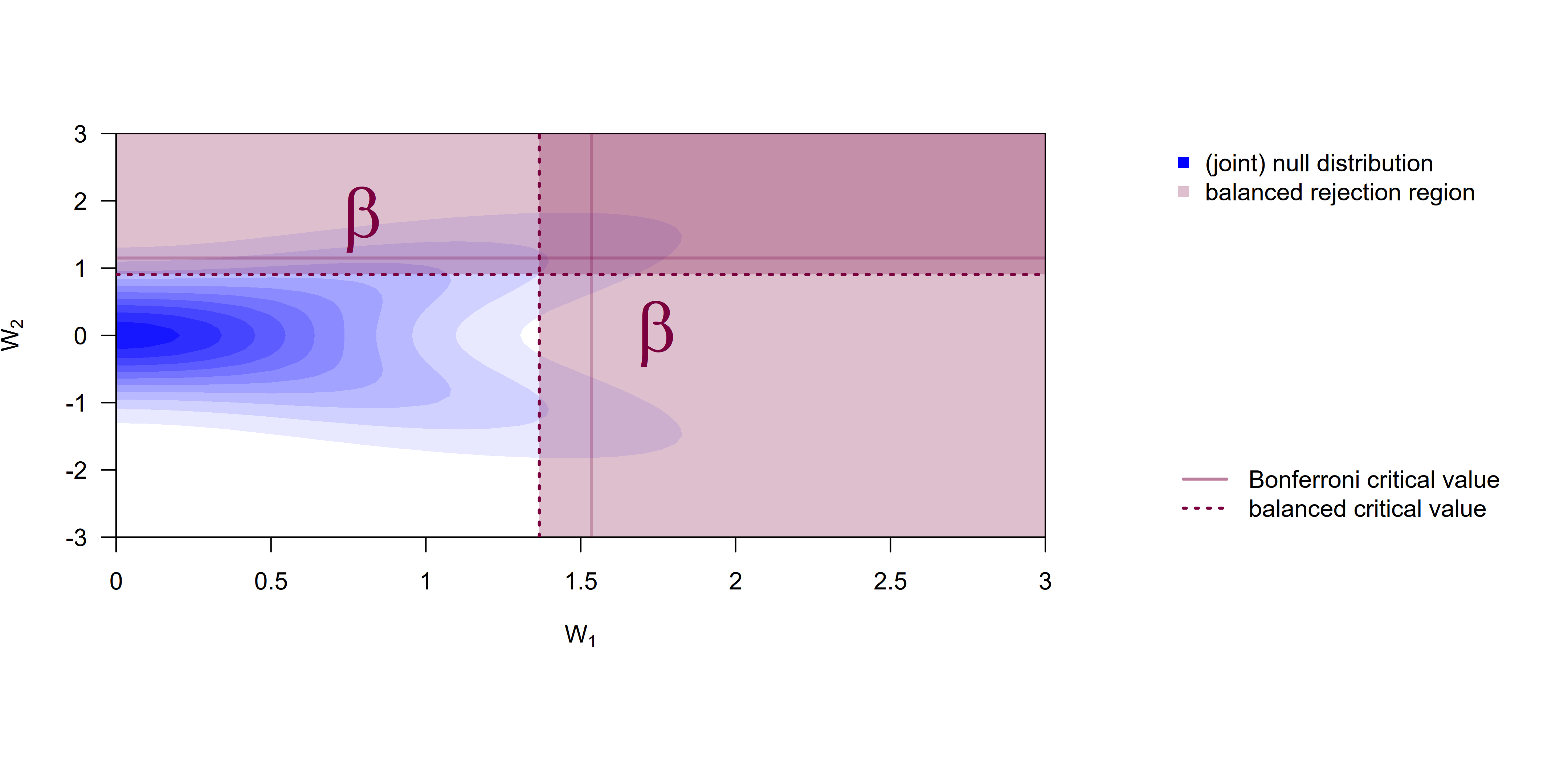}
    \caption{Exemplary illustration of the proposed balanced multiple testing procedure (dotted lines) for two test statistics with different distributions compared to the Bonferroni correction (solid lines).
    Integrating over the rejection regions for the first (right part) and second (upper part) hypothesis yields the same value $\beta$. }
    \label{fig:balanced}
\end{figure}

\textbf{Construction of balanced critical values}
If the joint cumulative distribution function $\mathcal F$ of the limit variable $(W_1,\ldots,W_L)$ is known, the critical values can be determined as $(1-\beta)$-quantiles of the marginal distributions, that are $q_{\ell,n} = \mathcal F_{\ell}^{-1}(1-\beta)$. Additionally, $\beta$ is chosen such that the family wise-error rate
\begin{align}
    \label{eq:betaF}
    1 - \mathcal F(\mathcal F_{1}^{-1}(1-\beta),\ldots,\mathcal F_{L}^{-1}(1-\beta)) \leq \alpha
\end{align}
is bounded by the global level $\alpha$.
%
It should be noted that, under Assumption~\ref{Wass}, $\beta$ 
can always be chosen as the Bonferroni-corrected local level $\alpha/L$, cf. \cite{dunn_multiple_1961}. An exemplary illustration of the proposed balanced multiple testing procedure compared to the Bonferroni correction can be found in Figure~\ref{fig:balanced}.
It is observable that larger values for $\beta$ generally decrease the critical values and, thus, increase the power of the multiple testing procedure.

To obtain maximal power, we aim to choose $\beta$ as the largest value that satisfies \eqref{eq:betaF}, that is
$$ \beta = \sup \{ \beta' \in [0,1] \mid 1 - \mathcal F(\mathcal F_{1}^{-1}(1-\beta'),\ldots,\mathcal F_{L}^{-1}(1-\beta')) \leq \alpha \}. $$
Throughout this paper, we consider the {upper} $p$-quantiles $F^{-1}(p) := \sup\{x \in\R \mid F(x) \leq p\}$ of a monotone non-decreasing function $F:\R \to [0,1]$.
Considering upper quantiles has the advantage
that the supremum is attained even for non-continuous cumulative distribution functions
compared to the lower (or any other) quantile. Hence, the chosen local level $\beta$ fulfills \eqref{eq:betaF} for non-continuous cumulative distribution functions as well; see also Remark~\ref{LowerQuant} in the appendix.


The preceding considerations motivate the following definition.
\begin{defi}[Asymptotic FWER]\label{aFWER}
Under the notation of Assumption~\ref{Wass},
    the asymptotic family-wise error rate $\mathrm{FWER}:\R \to [0,1]$ is defined by $$\mathrm{FWER}(\zeta) := 1- \mathcal F\left( \mathcal F_{1}^{-1}(1-\zeta),\dots,\mathcal F_{L}^{-1}(1-\zeta) \right) = P\left(\exists\ell\in\{1,\ldots,L\}:\: W_\ell > \mathcal F_{\ell}^{-1}(1-\zeta)\right)$$ for all $\zeta\in\R$, where here and throughout $F^{-1}(p) := \sup\{x \in\R \mid F(x) \leq p\}$ denotes the {upper} $p$-quantile, $p\in \mathbb R$, of a monotone non-decreasing function $F:\R \to [0,1]$. 
%
\end{defi}

Under Assumption~\ref{Wass}, we note that $\mathrm{FWER}$ is a monotone and continuous function.

In many applications, the joint distribution of $\left(W_\ell\right)_{\ell\in\{1,\ldots,L\}}$ depends on unknown parameters, cf. Examples~\ref{myEx}--\ref{myEx2}, or is approximated by a Monte-Carlo method due to the complexity of the distribution and quantile functions.
Even the marginal distributions may depend on unknown parameters, as, for instance, in Example~\ref{myEx} with unstudentized test statistics and in Example~\ref{myEx2}.
Hence, the cumulative distribution functions are usually approximated by (random) sequences $F_n$ and $F_{\ell,n}, \ell\in\{1,\ldots,L\}$ of cumulative distribution functions in practice.

\addtocounter{ex}{-2}
\begin{ex}[continued]
    By using the unstudentized test statistics in \eqref{eq:unstud}, the limit distribution depends on the covariance matrix $\boldsymbol{\Sigma}$ given in \eqref{eq:covmat}. The covariance matrix is usually unknown. However, it can be estimated, e.g., by $$\widehat{\boldsymbol{\Sigma}} := \frac{n}{n_1} S_1^2 \cdot \mathbf{1}_{(k-1)\times (k-1)} + \mathrm{diag}\left( \frac{n}{n_2} S_2^2, \ldots, \frac{n}{n_k} S_k^2  \right).$$
    Then, the joint cumulative distribution function $\mathcal F$ of $(| Z_1|, \ldots, | Z_{k-1}|)$ can be approximated by the cumulative distribution function $F_n$ of $(|\widehat Z_1|, \ldots, |\widehat Z_{k-1}|)$, where $(\widehat Z_1, \ldots, \widehat Z_{k-1})^\top := \widehat{\boldsymbol{\Sigma}}^{1/2} \mathbf{Y}$ and  $\mathbf{Y} \sim \mathcal{N}_{k-1} (\mathbf{0}, \mathbf{I})$ is a centered $(k-1)$-variate standard normal vector that is independent of the data $X_{11},\ldots,X_{kn_k}$.
    Moreover, the cumulative distribution function $\mathcal F_{\ell}$ of $|Z_\ell|$ can be approximated by the cumulative distribution function $F_{\ell,n}$ of $|\widehat Z_\ell|$ for all $\ell \in \{1,\ldots,k-1\}$.

    Alternatively, the cumulative distribution functions $F$ and $F_\ell$ can also be approximated by the empirical Monte-Carlo versions, that are
    \begin{align*}
        F_n(t_1,\ldots,t_{k-1}) = B_n^{-1} \sum_{b=1}^{B_n} \mathbbm{1}\{ |\widehat Z_{1;b}| \leq t_1, \ldots, |\widehat Z_{k-1;b}| \leq t_{k-1} \}\quad\text{and}\quad F_{\ell,n}(t) = B_n^{-1} \sum_{b=1}^{B_n} \mathbbm{1}\{ |\widehat Z_{\ell;b}| \leq t \}
    \end{align*}
     for $t_1,\ldots,t_{k-1},t\in\R$,
where $\mathbbm{1}$ denotes the indicator function, $B_n \in\mathbb N$ is the number of Monte-Carlo iterations and $(\widehat Z_{1;b}, \ldots, \widehat Z_{k-1;b})^\top := \widehat{\boldsymbol{\Sigma}}^{1/2} \mathbf{Y}_b$ for all $b\in\{1,\ldots,B_n\}$ with  $\mathbf{Y}_1,\ldots,\mathbf{Y}_{B_n} \sim \mathcal{N}_{k-1} (\mathbf{0}, \mathbf{I})$ being i.i.d. centered $(k-1)$-variate standard normal vectors that are independent of the data.    
\end{ex}\addtocounter{ex}{+1}

Similarly, the joint limit distribution for the Welch t-test statistics in Example~\ref{myEx} as well as the joint and marginal limit distributions of the test statistics in Example~\ref{myEx2} can be approximated. 

Of course, other methods for approximating the distribution functions are also possible.
For example, resampling procedures are often used to improve the small sample performance of the tests \cite{munko2023rmstbased, munko2025FMANOVA, munkoRMTL, Munko2025AStA, munko2026covariancecorrectionpermutationstatistics, baumeister2025early, sattler2025quadraticformbasedmultiple}, see also Section~\ref{sec:Consistent}. 
The applied approximations shall be maps defined on a suitable product probability space, which is formalized in the following assumption.
\begin{ass}\label{Fnass}
Let $(\Omega \times \Omega', \mathcal A \otimes \mathcal A', P \otimes P')$ denote a product probability space.
    For all $n\in\mathbb N$, we assume that $(F_n(\mathbf{t}))_{\mathbf t \in\R^L}: \Omega \times \Omega' 
    \to \mathrm{CDF}(\R^L)$ and
    $(F_{\ell,n}(t))_{t \in\R}: \Omega \times \Omega' 
    \to \mathrm{CDF}(\R), \ell\in\{1,\ldots,L\}.$
\end{ass}
The preceding assumption simply states that the approximations of the distribution functions should be random elements taking values in the suitable sets of cumulative distribution functions.
The random elements are defined on a product probability space $(\Omega \times \Omega', \mathcal A \otimes \mathcal A', P \otimes P')$. Here, $(\Omega, \mathcal A, P)$ describes the randomness from the data (as before) for a probability measure $P \in\mathcal P$, and $(\Omega', \mathcal A', P')$ describes additional introduced randomness, e.g., from resampling mechanisms.

Now, we can define suitable critical values based on the approximated cumulative distribution functions.
\begin{defi}[Approximated FWER and balanced critical values]\label{cv}
Under Assumption~\ref{Fnass}, let $\alpha \in (0,1)$ and define the approximated family-wise error rate by
    $$\mathrm{FWER}_{n}(\zeta) :=  1 - F_n\left( F_{1,n}^{-1}(1-\zeta),\dots,F_{L,n}^{-1}(1-\zeta)  \right)      \quad\text{ for all $\zeta\in \R, n\in\mathbb N$},$$ which takes values in the space of monotone, left-continuous functions $\R \to [0,1]$.
    Additionally, let $\beta_n := \mathrm{FWER}_{n}^{-1}(\alpha)$ be a sequence of maps $\Omega\times \Omega' \to [0,1]$.
    For $\ell\in\{1,\ldots,L\}$, we define the $\ell$th approximated balanced critical value with respect to $\alpha, F_n, F_{n,1},\ldots,F_{n,L}$ by
    $$q_{\ell,n} := F_{\ell,n}^{-1}\left( 1-\beta_n \right).$$ 
\end{defi}

\begin{figure}[bth]
    \centering
    \includegraphics[trim={0 0.65cm 1cm 1.5cm},
    clip, width=\linewidth]{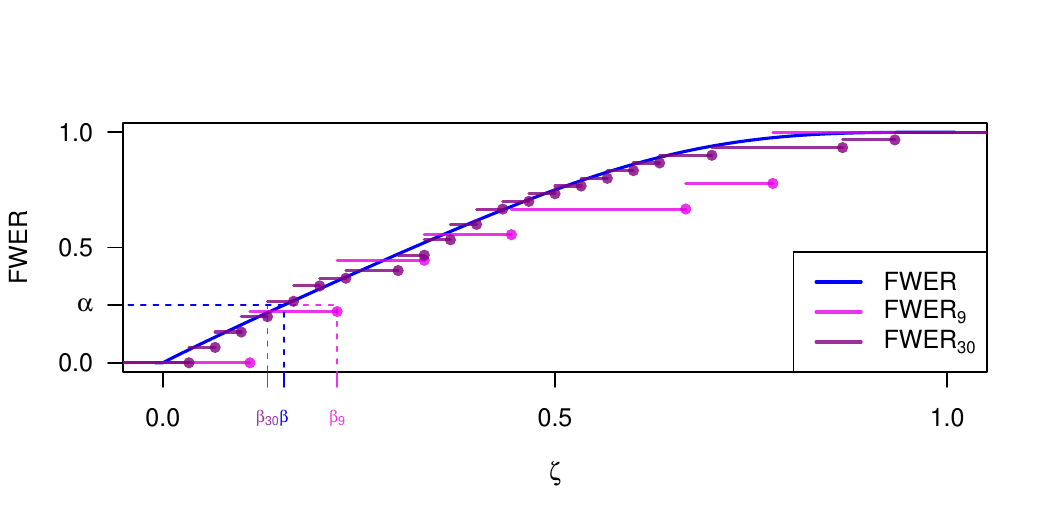}
    \caption{Exemplary illustration of a FWER curve and realizations of its approximation with $n \in \{9,30\}$. The approximations are calculated based on $B_n = n$ groupwise nonparametric bootstrap samples.}
    \label{fig:fwer}
\end{figure}

An exemplary illustration of an FWER curve 
as well as realizations for two approximations $\mathrm{FWER}_n$ with $n \in \{9,30\}$ is given in Figure~\ref{fig:fwer}. In addition, the approximations $\beta_n$ of the local level $\beta$ for the global level $\alpha = 0.25$ are visualized.


\section{Main results}\label{sec:Main}
In this section, we present the main theorem of this work, which states that the approximated balanced critical values $q_{1,n},\ldots,q_{L,n}$  with respect to  $\alpha, F_n, F_{1,n},\ldots,F_{L,n}$ from Definition~\ref{cv} lead to asymptotically balanced multiple tests that control the FWER in the strong sense as long as the approximations converge in outer probability to the true cumulative distribution functions.

\begin{theorem}\label{BSQuant}
Let $\alpha\in (0,1)$ and Assumptions~\ref{Wass} and~\ref{Fnass} hold.
Moreover, let $\mathrm{FWER}, q_{1,n}, \ldots, q_{L,n}$ be as in Definitions~\ref{aFWER} and~\ref{cv} and assume that $\mathrm{FWER}$ is strictly increasing on a compact interval $[a,b]$  with $a < \alpha/L \leq \alpha < b$. Furthermore, let
 \begin{align}
        \label{eq:condition}
        F_n(\mathbf{t}) \xrightarrow{P\otimes P'} \mathcal F(\mathbf{t}) \text{ as } n\to\infty \text{ for all } \mathbf{t}\in\R^L
\end{align}
    and 
\begin{align}
        \label{eq:condition2}
        F_{\ell,n}(t) \xrightarrow{P\otimes P'} \mathcal F_\ell (t) \text{ as } n\to\infty \text{ for all } t\in\R, \ell\in\{1,\ldots,L\}
    \end{align} hold for 
    $P \in \bigcup_{\ell=1}^L \mathcal P_\ell$.
    Then, it follows
     \begin{align}\label{eq:(i')}
         &\lim_{n\to\infty} \P^*\left(\exists \ell\in\T :\: W_{\ell,n} >  q_{\ell,n}\right) \leq \alpha
         &\quad\text{under $\bigcap_{\ell\in\T} \mathcal{H}_{0,\ell}$ for all index sets $\T\subset \{1,\dots,L\}$,}
         \\&\lim_{n\to\infty} \P^*\left(\exists \ell\in\{1,\dots,L\} :\: W_{\ell,n} >  q_{\ell,n}\right) = \alpha
         &\quad\text{under the global null hypothesis $\mathcal{H}_{0}$, and}\label{eq:(i'')}
         \\&\lim_{n\to\infty}\P^*\left(W_{\ell,n} >  q_{\ell,n}\right) = \mathrm{FWER}^{-1}(\alpha) 
         &\quad\text{under $\mathcal{H}_{0,\ell}$ for all $\ell\in\{1,\dots,L\}$}\label{eq:(ii')}.
     \end{align}
\end{theorem}
The preceding theorem provides that the multiple tests with the balanced critical values defined in Definition~\ref{cv} control the FWER in the strong sense by \eqref{eq:(i')} and fulfill the balancedness condition by \eqref{eq:(ii')}.
Here, it should be noted that \eqref{eq:(i)} and \eqref{eq:(ii)} are direct consequences from \eqref{eq:(i')} and \eqref{eq:(ii')}.

Furthermore, note that the conditions \eqref{eq:condition} and \eqref{eq:condition2} are required to hold if at least one of the local null hypotheses is true. This means that the approximations need to converge in outer probability to the true cumulative distribution functions even if some of the null hypotheses are false.
This is of course satisfied if the conditions \eqref{eq:condition} and \eqref{eq:condition2} hold for all probability measures $P$ in the considered class $\mathcal P$.

\begin{rem}[Comparison to the result of \cite{beran1988}]
The statements \eqref{eq:(i'')} and \eqref{eq:(ii')} are similar to those of Theorem~4.1 in \cite{beran1988}. 
However, the setups differ in some points:
As mentioned earlier, our method does not require measurability of the quantities.
Moreover, we do not require that $F_{1,n},\ldots,F_{L,n}$ are the marginals of $F_n$ but may rely on different approximation mechanisms, which may be useful, e.g., when the marginal distributions are known but the joint distribution is not, cf. Example~\ref{myEx} with Welch t-test statistics.
On the other hand, Theorem~4.1 in \cite{beran1988} covers the case of testing infinitely many hypotheses simultaneously.

Another difference is the probability measures on which the convergence results are required (and provided):
We require \eqref{eq:DistrConv} for all probability measures that suffices $\bigcap_{\ell \in\mathcal T} \mathcal{H}_{0,\ell}$ for all $\mathcal T \subset \{1,\ldots,L\}$ and \eqref{eq:condition} and \eqref{eq:condition2} for all probability measures $P \in \bigcup_{\ell=1}^L \mathcal P_\ell$.
In contrast, \cite{beran1988} requires the distributional convergence (their Assumption~2) and the convegence in probability (their Assumption~1) in $P_{\theta_n,n}$-probability for all sequences $(\theta_n)_{n\in\mathbb N}$ with $\theta_n \to \theta$ as $n\to\infty$, see \cite{beran1988} for further details. 
Technically, this seems to be harder to prove in many applications since convergence results over changing probability measures are more unusual than convergence results over a fixed probability measure.
\end{rem}

Based on the results of Theorem~\ref{BSQuant}, the decision rules for the global and local hypotheses are constructed as follows:
\begin{itemize}
	\item For each $\ell\in\{1,\ldots,L\}$, we reject the $\ell$th local hypothesis $\mathcal H_{0,\ell}$ if and only if $W_{\ell,n}>q_{\ell,n}$.
	\item We reject the global null hypothesis $\mathcal H_0$ whenever at least one of the hypotheses $\mathcal H_{0,1},\ldots,\mathcal H_{0,L}$ is rejected. Hence, we reject the global null hypothesis $\mathcal H_0$ if and only if
		there exist an $\ell\in\{1,\ldots,L\}$ such that $\ {W_{\ell,n}} > {q_{\ell,n}} .$

\end{itemize}
Here, each test statistic $W_{\ell,n}, \ell\in\{1,\ldots,L\},$ is treated in the same way and has asymptotically the same impact since we use the same local level of significance ${\beta}_n$ for each local hypothesis, cf. Definition~\ref{cv}.
The resulting tests can be formulated accordingly as $\varphi_\ell := \mathbbm{1}\{W_{\ell,n}>q_{\ell,n}\}$ for $\mathcal H_{0,\ell}$ for all $\ell\in\{1,\ldots,L\}$ and as $$\varphi := \mathbbm{1}\{\exists \,{\ell \in\{ 1, \ldots,L\}}: \ {W_{\ell,n}} > {q_{\ell,n}} \} = \max_{\ell \in \{1,\ldots,L\}} \varphi_\ell$$ for $\mathcal H_{0}$.
Then, \eqref{eq:(i')}--\eqref{eq:(ii')} can be formulated equivalently as
\begin{align}\label{eq:(start)}
         &\lim_{n\to\infty} \E^*\left(\max_{\ell\in\T} \varphi_\ell\right) \leq \alpha
         &\quad\text{under $\bigcap_{\ell\in\T} \mathcal{H}_{0,\ell}$ for all index sets $\T\subset \{1,\dots,L\}$,}
         \\&\lim_{n\to\infty} \E^*\left( \varphi\right) = \alpha
         &\quad\text{under the global null hypothesis $\mathcal{H}_{0}$, and}\label{eq:(mid)}
         \\&\lim_{n\to\infty}\E^*\left(\varphi_\ell\right) = \mathrm{FWER}^{-1}(\alpha)  
         &\quad\text{under $\mathcal{H}_{0,\ell}$ for all $\ell\in\{1,\dots,L\}$},\label{eq:(end)}
     \end{align}
where $\E^*$ denotes the outer expectation with respect to $(\Omega \times \Omega', \mathcal A \otimes \mathcal A', P \otimes P')$.

In the following remark, we give a requirement under which condition \eqref{eq:condition2} easily follows from condition \eqref{eq:condition}.

\begin{rem}\label{Remark}
    If $F_{1,n},\dots,F_{L,n}$ in Theorem~\ref{BSQuant} are the marginal cumulative distribution functions of $F_n$, \eqref{eq:condition2} is a direct consequence of \eqref{eq:condition}.
\end{rem}

It is well known that the closed testing procedure may improve the power of multiple tests. Hence, we propose a stepwise extension of the multiple testing procedure by applying the closed testing procedure in the following remark. 

\begin{rem}[Stepwise Extension]\label{stepwise}
Our methodologies can be combined with the closed testing procedure as in \cite{stepwise} to gain more power. 
Therefore,
for each $\ell\in\{1,\ldots,L\}$, the hypothesis $\mathcal{H}_{0,\ell}$ is rejected at level $\alpha$ if and only if for each $\mathcal{J}\subset\{1,\ldots,L\}$ with $\mathcal J \ni \ell$ the intersection hypothesis $\mathcal{H}_{0,\mathcal{J}} := \bigcap_{j\in\mathcal{J}} \mathcal{H}_{0,j}$ is rejected at level $\alpha$. For testing an intersection hypothesis  $\mathcal{H}_{0,\mathcal{J}}$, we can use the procedure as described above. 
Further details on this matter can be found in Appendix~\ref{sec:ProofStepwise}.
The resulting tests control the family-wise error rate in the strong sense, while the local asymptotic power may be increased compared to the tests from Theorem~\ref{BSQuant}.
However, the tests generally do not satisfy the balancedness condition \eqref{eq:(ii)}.
\end{rem}

In many applications, hypotheses are formulated in terms of statistical functionals of the probability measure, also called estimands. Examples for estimands are location parameters like the mean \cite{Baumeister2025} and quantiles \cite{baumeister2024, baumeister2025early}, dispersion parameters like covariance and correlation matrices \cite{sattler2022,sattler2023} as well as
estimands from survival analysis like the restricted mean survival time and the restricted mean time lost \cite{munko2023rmstbased,munko2025effect,munkoRMTL}.
If multiple tests for estimands are present, simultaneous confidence regions for the estimands can be obtained. This is explained in more detail in the following remark. 

\begin{rem}[Simultaneous Confidence Regions]\label{confidence}
    Let us consider classes of local hypotheses of the form
$\mathcal H_{0,\ell}: {h}_\ell(\P) = {\xi}_\ell$ for ${\xi}_\ell\in\Xi_\ell, \ell\in\{1,\ldots,L\},$
about estimands ${h}_\ell: \mathcal P \to \Xi_\ell$ for all $\ell\in\{1,\ldots,L\}$, where $\Xi_1,\ldots,\Xi_L$ denote arbitrary sets (e.g., $\Xi_\ell = \R$).
Moreover, we write $W_{\ell,n} = W_{\ell,n}({\xi}_\ell)$ for the $\ell$th local test statistic and $q_{\ell,n} = q_{\ell,n}(\xi_\ell)$ for the critical value to express their dependence on ${\xi}_\ell$. 
   Then, we can use the constructed multiple testing procedure to define simultaneous balanced confidence regions for the estimands ${h}_1(\P),\ldots,{h}_L(\P)$ with asymptotic global confidence level $1-\alpha$. Under the notation and conditions of Theorem~\ref{BSQuant}, we define the $\ell$th confidence region as
   $$ CR_{\ell,n} := \left\{ \xi \in \Xi_{{\ell}} \mid   W_{\ell,n}(\xi) \leq q_{\ell,n}(\xi) \right\} $$ for all $\ell\in\{1,\ldots,L\}$.
Then, it can be easily checked that
$\lim_{n\to\infty}\P^*(\exists\ell\in\{1,\dots,L\}:\: {h}_\ell(\P) \not\in CR_{\ell,n} ) = \alpha$ holds for all $P\in\mathcal P$.
Note that the confidence region depends on the confidence level $1-\alpha$ through the critical value, cf. Definition~\ref{cv}.

By Remark~\ref{stepwise} and Appendix~\ref{sec:ProofStepwise}, possibly smaller but usually more complicated confidence regions are given by
$$ CR_{\ell,n} := \left\{ \xi \in \Xi_{{\ell}} \mid  \exists \mathcal{J}\subset \{1,\ldots,L\} \text{ with } \ell \in \mathcal{J}: \:  \forall \, {j \in\mathcal{J}}:\: {W_{j,n}(\xi)}\leq q_{j,n}^{\mathcal J}(\xi) \right\} ,\quad \ell\in\{1,\ldots,L\}.$$

\end{rem}

For instance, the previous remark yields tools for constructing simultaneous balanced confidence intervals for the mean differences $\mu_{\ell+1} - \mu_1, \ell\in\{1,\ldots,k-1\}$ in an adaptation of Example~\ref{myEx}.

\begin{ex}
    Let us consider the situation of Example~\ref {myEx} again with the following adaptation:
    We consider tests for the class of local hypotheses $\mathcal H_{0,\ell}: \mu_{\ell +1} - \mu_1 = \xi_\ell$ for $\xi_\ell \in \mathbb R, \ell\in\{1,\ldots,k-1\}$.
    Suitable test statistics are, for example, the following unstudentized test statistics
    \begin{align*}
        W_{\ell,n}(\xi_\ell) = \sqrt{n}{|\bar{X}_{\ell + 1} - \bar{X}_{1} - \xi_\ell|}, \quad \ell\in\{1,\ldots,k-1\},
    \end{align*}
    with the same limit distributions under the null hypotheses and the same approximations of the cumulative distribution functions as previously described in Example~\ref{myEx}.
    
    By the preceding remark, simultaneous asymptotic confidence intervals for the estimands $\mu_2 - \mu_1, \ldots, \mu_k - \mu_1$ are given by 
    \begin{align*}
        CR_{\ell,n} & = \left\{ \xi \in \mathbb R \mid   W_{\ell,n}(\xi) \leq q_{\ell,n} \right\}
        = \left[ \bar{X}_{\ell + 1} - \bar{X}_{1} - \frac{q_{\ell,n}}{\sqrt{n}},  \bar{X}_{\ell + 1} - \bar{X}_{1} + \frac{q_{\ell,n}}{\sqrt{n}} \right], \quad \ell\in\{1,\ldots,k-1\},
    \end{align*}
    where $q_{1,n},\ldots,q_{L,n}$ are the balanced critical values from Definition~\ref{cv}.
\end{ex}

\section{Consistent Resampling Schemes}\label{sec:Consistent}
In many scenarios, resampling schemes like bootstrap or permutation approaches have been proven useful to approximate the limit distribution of statistics.
 In this section, we use that the consistency of a resampling scheme implies \eqref{eq:condition} for constructing resampling-based balanced multiple tests by applying Theorem~\ref{BSQuant} and construct adjusted p-values that lead to the same test decisions.

As the definition of a consistent resampling scheme is rather technical, we have moved the detailed description to Appendix~\ref{sec:ConsDetails}.

Lemma~\ref{ResamplingLemma} in Appendix~\ref{sec:ConsDetails} ensures that the conditional cumulative distribution function of a resampled test statistic 
from a consistent resampling scheme
fulfills \eqref{eq:condition}.
In practice, the conditional distribution of resampling statistics is often approximated through a Monte Carlo method. 
Lemma~\ref{MonteCarloRemark} in Appendix~\ref{sec:ConsDetails} addresses this circumstance: It provides that the empirical cumulative distribution function $F_n$ of $B_n$ independently resampled test statistics $({W}_{1,n}^{(b)},\dots,{W}_{L,n}^{(b)})$, $b\in\{1,\ldots,B_n\},$ from a consistent resampling scheme
fulfills the condition \eqref{eq:condition} to approximate $\mathcal F$ in Theorem~\ref{BSQuant} if $B_n \to \infty$ as $n\to\infty$.
For that choice of $F_n$, we obtain
\begin{align*}
    \mathrm{FWER}_{n}(\zeta) = \frac{1}{B_n} \sum\limits_{b=1}^{B_n} \mathbbm{1}\left\{ \exists \ell\in\{ 1,\ldots,L \} : {W}_{\ell,n}^{(b)} > F_{\ell,n}^{-1}(1-\zeta) \right\}
\end{align*} for all $\zeta\in \R.$
Regarding Remark~\ref{Remark}, the marginal cumulative distribution functions $\mathcal F_\ell$ can be approximated by the marginals of the empirical cumulative distribution function $F_n$ that are given by
\begin{align}
    \label{eq:myFnl}
     F_{\ell,n}(w) := \frac{1}{B_n}\sum_{b=1}^{B_n} \mathbbm{1}\left\{ {W}_{\ell,n}^{(b)} \leq w\right\}, \quad \text{for all }\ell\in\{1,\dots,L\}, w\in\R.
\end{align}
 Then, the local level ${\beta}_n$ is the largest value such that the approximated family-wise error rate is bounded by the level of significance $\alpha$, i.e.,
\begin{align*}
    {\beta}_n = \mathrm{FWER}_{n}^{-1}(\alpha) &= \max\left\{ \zeta \in \left\{ \frac{1}{B_n}, \frac{2}{B_n}, \ldots, 1  \right\} \mid  \mathrm{FWER}_{n}(\zeta) \leq \alpha \right\}
\end{align*}
Note that we only have to consider $\zeta \in \left\{ \frac{1}{B_n}, \frac{2}{B_n}, \ldots, 1 \right\}$ since the quantiles can only take $B_n$ different values, respectively.
Additionally, we have ${\beta}_n \geq \frac{1}{B_n}\left\lfloor\frac{B_n\alpha}{L}\right\rfloor$.
The lower bound can be interpreted as Bonferroni bound and results from the following inequalities:
\begin{align*}
    \mathrm{FWER}_{n}\left(\frac{1}{B_n}\left\lfloor\frac{B_n\alpha}{L}\right\rfloor\right)
    &\leq \sum\limits_{\ell=1}^L  \frac{1}{B_n} \sum\limits_{b=1}^{B_n}\mathbbm{1}\left\{  {W}_{\ell,n}^{(b)} > F_{\ell,n}^{-1}\left(1-\frac{1}{B_n}\left\lfloor\frac{B_n\alpha}{L}\right\rfloor\right) \right\}
    \leq L \frac{1}{B_n}\left\lfloor\frac{B_n\alpha}{L}\right\rfloor \leq \alpha.
\end{align*}

With the above choices for $F_n, F_{\ell,n}$ and $\beta_n$, Theorem~\ref{BSQuant} implies that
$q_{\ell,n} = F_{\ell,n}^{-1}(1-\beta_n),\ell\in\{1,\ldots,L\},$ fulfills \eqref{eq:(i')}--\eqref{eq:(ii')}
under the conditions of Theorem~\ref{BSQuant} and Lemma~\ref{MonteCarloRemark}.

\begin{cor}\label{Cor}
    Let $\alpha\in (0,1)$, Assumptions~\ref{Wass}--\ref{Fnass} hold, and $B_n\to\infty$ as $n\to\infty$.
Moreover, let $\mathrm{FWER}$ be as in Definition~\ref{aFWER} and assume that $\mathrm{FWER}$ is strictly increasing on $[a,b]$  with $a < \alpha/L \leq \alpha < b$.
 Furthermore, let $F_n$ be the empirical cumulative distribution function of $B_n$ independently resampled test statistics of a resampling scheme that is consistent for $(W_1,\ldots,W_L)$, see Appendix~\ref{sec:ConsDetails} for details.
 If $q_{1,n},\ldots,q_{L,n}$ are the critical values as in Definition~\ref{cv} with respect to $\alpha$, $F_n$, and $F_{1,n},\ldots,F_{L,n}$ in \eqref{eq:myFnl}, it follows \eqref{eq:(i')}--\eqref{eq:(ii')}.
\end{cor}

\textbf{Adjusted p-values}
   The method described in Corollary~\ref{Cor} for constructing balanced multiple tests based on a consistent resampling scheme is accompanied by an adjustment of p-values. To see this, we define the local p-value for the $\ell$th hypothesis $\mathcal H_{0,\ell}$ by
   $$ \beta_{\ell,n} := 
  \frac{1}{B_n}\left( 1 + \sum_{b=1}^{B_n} \mathbbm{1}\left\{ W_{\ell,n}^{(b)} \geq W_{\ell,n}  \right\}\right)
   =1 - F_{\ell,n -}(W_{\ell,n}) + B_n^{-1}$$
    for all $\ell\in\{1,\ldots,L\}$.
    Here and throughout, $F_-$ denotes the left-continuous version of a monotone, right-continuous function $F$.
   Comparing the local p-values to $\beta_n$ yields multiple test decisions that are consistent with the method in Theorem~\ref{BSQuant}. 
   Translating this comparison to a comparison with the level of significance $\alpha$ is intuitive due to the definition of $\beta_n = \mathrm{FWER}_n^{-1}(\alpha)$.
   Hence, by plugging the local p-value in $\mathrm{FWER}_{n}$, the adjusted p-value for the $\ell$th hypothesis $\mathcal H_{0,\ell}$ can be defined by
   $$p_{\ell} := \mathrm{FWER}_{n}\left(\beta_{\ell,n}\right) $$ for all $\ell\in\{1,\ldots,L\}$ and the global p-value for $\mathcal H_0$ by $p := \min\{p_1,\ldots,p_L\}$. 
   The following proposition ensures that the test decisions based on these p-values are unchanged.
   \begin{theorem}
       \label{pValues}
    With the notation of Section~\ref{sec:Consistent},
       \begin{enumerate}
       \item[(1)] for each $\ell\in\{1,\ldots,L\}$, it holds $p_{\ell} \leq \alpha$ if and only if $W_{\ell,n} > q_{\ell,n}$,
       \item[(2)] it holds $p \leq \alpha$ if and only if there exist an $\ell\in\{1,\ldots,L\}$ such that ${W_{\ell,n}}>q_{\ell,n} $.
   \end{enumerate}
   \end{theorem}

\section{Discussion}\label{sec:Discussion}
In quantitative research, often more than one null hypothesis shall be tested. The most popular adjustment methods for controlling the family-wise error rate in multiple testing problems are the Bonferroni and the Holm adjustment.
However, they do not take the joint distribution of the test statistics into account, although this can yield considerably more power.
So far, the joint distribution of the test statistics has only been used in rather specific scenarios for multiple tests. In this work, we closed this gap by developing a flexible and general framework for constructing asymptotically balanced multiple tests. 
In addition, a stepwise extension as well as simultaneous balanced confidence regions were proposed.
Furthermore, we proved that the general framework can be applied if consistent resampling procedures are present. In that case, we additionally provided adjusted p-values that yield the same test decisions as the balanced multiple tests.

Our setup is not tailored to specific hypotheses, parameters, models, or test statistics but treats a general framework that can be applied to several statistical problems. Thus, it provides a foundation for a wide range of future research opportunities through its application to concrete settings.

Further possible future work includes the construction of a general methodology that uses the joint distribution of test statistics for other controlling measures in multiple testing problems, such as the false discovery rate.

Moreover, balanced multiple tests for infinitely many hypotheses instead of a finite number $L$ of hypotheses, similarly to the setup in \cite{beran1988}, could be studied in future research.

Due to the theoretical nature of this work, it does not contain any simulation studies or real data examples. However, we refer to the scenarios in which the approach has already been successfully applied \cite{munko2023rmstbased, munko2025FMANOVA, munkoRMTL, Munko2025AStA, munko2026covariancecorrectionpermutationstatistics, sattler2025quadraticformbasedmultiple} for simulation results and data applications. In these works, it has already been observed that the balanced multiple testing approach that uses the joint distribution of the test statistics has a notably higher power than standard adjustment methods like the Bonferroni correction.


\section*{Acknowledgements}
M. Munko wants to thank her former affiliations Otto von Guericke University Magdeburg and RWTH Aachen University, where parts of this work have been done.
She also gratefully acknowledges the valuable discussions with Dennis Dobler, Marc Ditzhaus, Markus Pauly, and Marléne Baumeister, as well as Paavo Sattler for his insightful comments on the manuscript.

This work was supported by national funds through FCT,I.P. (Fundação para a Ciência e a Tecnologia,I.P.), under the project - UID/04152/2025 - Centro de Investigação em Gestão de Informação (MagIC)/NOVA IMS (DOI: 10.54499/UID/04152/2025).

\bibliographystyle{abbrvnat}
\bibliography{sample}

\appendix

\section{Details on resampling schemes}\label{sec:ConsDetails}
In this section, we formalize the resampling test statistic and define consistency of a resampling scheme. Moreover, we will conclude that \eqref{eq:condition} follows for the conditional cumulative distribution function of a consistent resampling scheme in Lemma~\ref{ResamplingLemma} and its Monte Carlo version in Lemma~\ref{MonteCarloRemark}.

Mathematically, we assume that a resampling scheme for a vector of $L$ test statistics is given through a map $$\mathbf{W}_n:\Omega \times \widetilde{\Omega} \to \R^L$$ from a product probability space $(\Omega \times \widetilde{\Omega}, \mathcal A \otimes \widetilde{\mathcal A}, P \otimes \widetilde{P})$, where $(\Omega, \mathcal A, P)$ describes the randomness from the data (as before) and $(\widetilde{\Omega}, \widetilde{\mathcal A}, \widetilde{P})$ describes the randomness introduced by the resampling method.
Note that the test statistics $W_{1,n},\ldots,W_{L,n}: \Omega  \to \mathbb R$ introduced in Section~\ref{sec:MultiSetup} are only defined on the first probability space.
Similarly, the hypotheses formulated in Section~\ref{sec:MultiSetup} concern only the first probability measure $P$.

A resampling scheme $\mathbf{W}_n:\Omega \times \widetilde{\Omega} \to \R^L$ is called \emph{consistent} for $(W_1,\dots,W_L)$ if $\mathbf{W}_n$ converges weakly conditionally on the data in outer probability to $(W_1,\dots,W_L)$, i.e., to the same limit distribution as the original test statistics $(W_{1,n},\ldots,W_{L,n})$ under the global null hypothesis, cf. Assumption~\ref{Wass}.
Conditionally weak convergence can be formalized as follows \cite{DeltaMathod,vaartWellner2023}:
\begin{align}\label{eq:precise}
    &\sup\limits_{h\in BL_1(\R^L)} \left| \E_2 \left[h\left(\mathbf W_n \right)^{2*}\right]  - \E \left[h(W_1,\ldots,W_L)\right] \right| \xrightarrow{P} 0\quad &\text{as $n\to\infty$},
       \\& \label{eq:precise2}
        \E_2\left[ h\left(\mathbf W_n \right)^* \right] - \E_2 \left[h\left( \mathbf W_n \right)_*\right]  \xrightarrow{P} 0\quad&\text{as $n\to\infty$ for all $h\in BL_1(\R^L)$}
    \end{align} for all $P \in\bigcup_{\ell\in\{1,\ldots,L\}}\mathcal P_\ell$,
    where $BL_1(\R^L)$ denotes the space of all Lipschitz functions $\R^L \to \R$ with Lipschitz norm bounded by $1$, $\E_2$ denotes the expectation with respect to $(\widetilde{\Omega},\widetilde{\mathcal A}, \widetilde{P})$, and the super- and subscript asterisks denote the minimal measurable majorants and maximal measurable minorants, respectively, with respect to $\Omega\times\widetilde{\Omega}$ jointly
    for $^*$ and with respect to $\widetilde{\Omega}$ for $^{2*}$, see \cite{vaartWellner2023} for details.

The following lemma states that the conditional cumulative distribution function of a resampled test statistic fulfills the condition to approximate the function $\mathcal F$ in Theorem~\ref{BSQuant} for a consistent resampling scheme.

\begin{lem}\label{ResamplingLemma}
    Let Assumption~\ref{Wass} hold.
    Suppose that $\mathbf W_n, n\in\mathbb{N},$ is a resampling scheme and
    \begin{align*}
        G_n(\mathbf{t}) := \E_2\left[\mathbbm{1}\left\{ \mathbf{W}_n \leq \mathbf{t}\right\}^{2*}\right]\quad\text{for all }\mathbf t \in\R^L
    \end{align*} is the conditional cumulative distribution function of the resampling statistic.
    Here,
    $\E_2$ denotes the expectation with respect to the probability space $(\widetilde{\Omega},\widetilde{\mathcal A}, \widetilde{P})$,
    $^{2*}$ denotes the minimal measurable majorant with respect to $\widetilde{\Omega}$,
    and $\mathbf{W}_n \leq \mathbf{t}$ means that all components of $\mathbf{W}_n$ are less than or equal to the corresponding values in the vector $\mathbf{t}$. 
    If $\mathbf W_n$ is consistent for $(W_1,\dots,W_L)$ for $P\in\bigcup_{\ell=1}^L \mathcal{P}_\ell$, condition \eqref{eq:condition} is satisfied for $G_n$.
\end{lem}
\begin{proof}[Proof of Lemma~\ref{ResamplingLemma}]
    Let $\mathbf{t}\in\R^L$ be arbitrary and $\mathbf W := (W_1,\ldots,W_L)$.
    Approximate $f:\R^L\to\{0,1\}, f(\mathbf{w}) = \mathbbm{1}\{\mathbf{w}\leq \mathbf{t}\} $ through sequences of Lipschitz functions $(g_m)_{m\in\mathbb N}, (h_m)_{m\in\mathbb N}$ with $1\geq g_m \geq f \geq h_m \geq 0$ and $\E\left[g_m(\mathbf{W}) - h_m(\mathbf{W})\right] \leq m^{-1}$ for all $m\in\mathbb{N}$, where $\mathbf{w} \leq \mathbf{t}$ means that all components of $\mathbf{w}$ are less than or equal to the corresponding values in the vector $\mathbf{t}$.
    Let $\E_2^*$ the outer expectation with respect to $(\widetilde{\Omega},\widetilde{\mathcal A}, \widetilde{P})$, cf. Section~1.2 in \cite{vaartWellner2023}.
    Then, it holds
    \begin{align*}
    \left|\E_2\left[ f(\mathbf{W}_n)^* \right] - \E\left[ f(\mathbf{W}) \right] \right| 
    &\leq \E_2\left[ f(\mathbf{W}_n)^* \right] - \E_2^*\left[ f(\mathbf{W}_n) \right] + \left|\E_2^*\left[ f(\mathbf{W}_n) \right] - \E\left[ f(\mathbf{W}) \right] \right|
    \\&\leq \E_2\left[ g_m(\mathbf{W}_n)^* \right] - \E_2^*\left[ h_m(\mathbf{W}_n) \right] \\&\quad    +\max\left\{ \E_2^*\left[ g_m(\mathbf{W}_n) \right] - \E\left[ g_m(\mathbf{W}) \right] ,   \E\left[ h_m(\mathbf{W}) \right] - \E_2^*\left[ h_m(\mathbf{W}_n) \right] \right\}
     \\&\quad    + \E\left[ g_m(\mathbf{W}) \right] - \E\left[ h_m(\mathbf{W}) \right]
    \\&\leq \E_2\left[ g_m(\mathbf{W}_n)^*\right] - \E_2\left[g_m(\mathbf{W}_n)_* \right] + \E_2^*\left[g_m(\mathbf{W}_n)\right] - \E_2^*\left[h_m(\mathbf{W}_n)\right]
        \\&\quad    +\max\left\{ \E_2^*\left[ g_m(\mathbf{W}_n) \right] - \E\left[ g_m(\mathbf{W}) \right] ,   \E\left[ h_m(\mathbf{W}) \right] - \E_2^*\left[ h_m(\mathbf{W}_n) \right] \right\} + m^{-1}
        \\&\xrightarrow{P} 0 + \E\left[g_m(\mathbf{W})\right] - \E\left[h_m(\mathbf{W})\right] + 0 + m^{-1} \leq 2m^{-1}
    \end{align*}
    as $n\to\infty$, where here and throughout $A^*$ and $A_*$ denote the minimal measurable majorant and maximal measurable minorant, respectively, for a map $A:\Omega\times\widetilde{\Omega} \to\mathbb R$ with respect to $\Omega\times\widetilde{\Omega}$ jointly.
    By choosing $m$ sufficiently large, it follows
    \begin{align}\label{eq:fmajor}
        \E_2\left[ f(\mathbf{W}_n)^* \right] \xrightarrow{P} \E\left[ f(\mathbf{W}) \right] = P(\mathbf W \leq \mathbf t) = \mathcal F(\mathbf{t})
    \end{align} as $n\to\infty$.
    Analogously, one can show
    \begin{align}\label{eq:fminor}
        \E_2\left[ f(\mathbf{W}_n)_* \right] \xrightarrow{P}  \mathcal F(\mathbf{t})
    \end{align}
         as $n\to\infty$. 
Hence, we have
     \begin{align*}
      \left| G_n(\mathbf{t}) - \mathcal F(\mathbf{t}) \right|^* 
      &= \left|\E_2\left[f( \mathbf{W}_n )^{2*}\right] - \mathcal F(\mathbf{t}) \right|^* 
      \\&\leq \left(\E\left[f( \mathbf{W}_n )^{*}\right]  - \E_2\left[f( \mathbf{W}_n )^{2*}\right]\right)^* +  \left|\E\left[f( \mathbf{W}_n )^{*}\right] - \mathcal F(\mathbf{t}) \right| 
      \\&\leq \E\left[f( \mathbf{W}_n )^{*}\right]  - \E_2\left[f( \mathbf{W}_n )_*\right] +  \left|\E\left[f( \mathbf{W}_n )^{*}\right] - \mathcal F(\mathbf{t}) \right|
      \\ &\xrightarrow{P} \mathcal F(\mathbf{t}) - \mathcal F(\mathbf{t}) + |\mathcal F(\mathbf{t})-\mathcal F(\mathbf{t})| = 0
    \end{align*}
     as $n\to\infty$. Thus, (\ref{eq:condition}) in Theorem~\ref{BSQuant} follows.
\end{proof}

Next, we turn to the Monte Carlo approximation of such a conditional cumulative distribution function. Hence, we aim to show that \eqref{eq:condition} in Theorem~\ref{BSQuant} follows for consistent resampling schemes,
where $F_n$ is the empirical cumulative distribution function of independently resampled test statistics.
To do so, we need to formalize the concept of independently resampled test statistics: Let $(\Omega, \mathcal A, P)$ again describe the randomness from the data and $(\Omega', \mathcal A', P') = (\widetilde{\Omega}^{\mathbb N}, \widetilde{\mathcal A}^{\otimes \mathbb N}, \widetilde{P}^{\otimes\mathbb N})$ describe the randomness introduced by independent resampling schemes.
Then, we call $$\mathbf{W}_n^{(b)}: \Omega\times \widetilde{\Omega}^{\mathbb N} \to \mathbb{R}^L,\quad  \mathbf{W}_n^{(b)}(\omega_0, \omega_1,\ldots) := \mathbf{W}_n(\omega_0,\omega_b), \qquad b\in\mathbb N,$$
\textit{independently resampled test statistics} of the resampling scheme $\mathbf{W}_n$.
We equip $\Omega\times \widetilde{\Omega}^{\mathbb N}$ with the product $\sigma$-algebra and product probability measure. 

The following lemma states that the empirical cumulative distribution function of an increasing number $B_n$ of independently resampled test statistics 
fulfills the condition to approximate the function $\mathcal F$ in Theorem~\ref{BSQuant} for a consistent resampling scheme.

\begin{lem}\label{MonteCarloRemark}
    Let Assumption~\ref{Wass} hold and $(B_n)_{n\in\mathbb N}\subset\mathbb N$ be a sequence with $B_n \to\infty$ as $n\to\infty$.
    Suppose that $F_n$ is the empirical cumulative distribution function of $B_n$ independently resampled test statistics $\mathbf{W}_n^{(1)},\ldots,\mathbf{W}_n^{(B_n)}$ of a resampling scheme $\mathbf W_n$ for all $n\in\mathbb{N}$, 
    i.e.,
    \begin{align*}
        F_n(\mathbf{t}) := \frac{1}{B_n}\sum_{b=1}^{B_n} \mathbbm{1}\left\{ \mathbf{W}_n^{(b)} \leq \mathbf{t}\right\}\quad\text{for all }\mathbf t \in\R^L.
    \end{align*}
    If $\mathbf W_n$ is consistent for $(W_1,\dots,W_L)$ for $P\in\bigcup_{\ell=1}^L \mathcal{P}_\ell$, condition \eqref{eq:condition} is satisfied for $F_n$.
\end{lem}
\begin{proof}[Proof of Lemma~\ref{MonteCarloRemark}]
    Let $\mathbf{t}\in\R^L$ be arbitrary, $\mathbf W := (W_1,\ldots,W_L)$, and $f:\R^L\to\{0,1\}, f(\mathbf{w})= \mathbbm{1}\{\mathbf{w}\leq \mathbf{t}\} $.
       Moreover, let $\E_2$ denote the expectation with respect to the product probability space $(\Omega', \mathcal A', P') = (\widetilde{\Omega}^{\mathbb N}, \widetilde{\mathcal A}^{\otimes \mathbb N}, \widetilde{P}^{\otimes\mathbb N})$ and $\E_2^*$ the outer expectation with respect to $(\Omega', \mathcal A', P')$.
    Since $B_n\to\infty$ as $n\to\infty$, we have
     \begin{align*}
      &\E_2\left[ \left(\left| F_n(\mathbf{t}) - \mathcal F(\mathbf{t}) \right|^2\right)^* \right]
        \\& \leq   \E_2\left[ \left(F_n^2(\mathbf{t})\right)^*\right] +2 \E_2\left[\left(-F_n(\mathbf{t})\right)^* \right]\mathcal F(\mathbf{t}) + \mathcal F^2(\mathbf{t})
        \\& \leq  \frac{1}{B_n^2}\sum_{b_1,b_2=1}^{B_n} \E_2\left[\left(f(\mathbf{W}_{n}^{(b_1)})f(\mathbf{W}_{n}^{(b_2)})\right)^*\right]  + \frac{2}{B_n}\sum_{b=1}^{B_n} \E_2\left[\left(-f(\mathbf{W}_{n}^{(b)})\right)^*\right]\mathcal F(\mathbf{t}) + \mathcal F^2(\mathbf{t}) 
       \\& \leq  \frac{1}{B_n} + \frac{1}{B_n^2}\sum_{b_1,b_2=1, b_1 \neq b_2}^{B_n} \E_2\left[f(\mathbf{W}_{n}^{(b_1)})^* \right]\E_2\left[f(\mathbf{W}_{n}^{(b_2)})^*\right]  - {2}\E_2\left[f(\mathbf{W}_{n}^{(1)})_*\right]\mathcal F(\mathbf{t}) + \mathcal F^2(\mathbf{t})
       \\& \leq  \frac{1}{B_n} + \frac{B_n -1}{B_n} \E_2\left[f(\mathbf{W}_{n}^{(1)})^* \right]^2  - {2}\E_2\left[f(\mathbf{W}_{n}^{(1)})_*\right]\mathcal F(\mathbf{t}) + \mathcal F^2(\mathbf{t})
        \\& \xrightarrow{P}  \mathcal F^2(\mathbf{t}) -2 \mathcal F(\mathbf{t})\mathcal F(\mathbf{t}) + \mathcal F^2(\mathbf{t}) = 0
    \end{align*}
     as $n\to\infty$ by \eqref{eq:fmajor} and \eqref{eq:fminor}. Thus, $\E^*\left[ \left| F_n(\mathbf{t}) - \mathcal F(\mathbf{t}) \right|^2 \right] \to 0$ as $n\to\infty$ by the dominated convergence theorem (dominated by $1$), where $\E^*$ denotes the outer expectation with respect to $(\Omega \times \Omega', \mathcal A \otimes \mathcal A', P \otimes P')$. Hence, (\ref{eq:condition}) is satisfied for $F_n$.
\end{proof}

\section{Details on the stepwise extension}\label{sec:ProofStepwise}
In this section, we give a more detailed description of the stepwise extension of the balanced multiple tests in Theorem~\ref{BSQuant} resulting from an application of the closed testing procedure.

Applying the closed testing procedure yields that, for each $\ell\in\{1,\ldots,L\}$, the hypothesis $\mathcal{H}_{0,\ell}$ is rejected at level $\alpha$ if and only if for each $\mathcal{J}\subset\{1,\ldots,L\}$ with $\mathcal J \ni \ell$ the intersection hypothesis $\mathcal{H}_{0,\mathcal{J}} := \bigcap_{j\in\mathcal{J}} \mathcal{H}_{0,j}$ is rejected at level $\alpha$. 
The intersection hypothesis $\mathcal{H}_{0,\mathcal{J}}$ can be rejected at level $\alpha$ whenever 
		${W_{j,n}} > q_{j,n}^{\mathcal J} := {F_{j,n}^{-1}\left( 1-\beta_n^{\mathcal J} \right)} $
 holds for some ${j \in\mathcal{J}}$, where
 $\beta_n^{\mathcal J}  := \mathrm{FWER}_{n}^{{\mathcal J},-1}(\alpha)$ 
     with  $$\mathrm{FWER}_{n}^{\mathcal J}(\zeta) :=  1 - F_n^{\mathcal J}\left( (F_{j,n}^{-1}(1-\zeta))_{j\in\mathcal J}  \right)      \quad\text{ for all $\zeta\in \R$}$$ and
    $F_n^{\mathcal J}$ denoting the marginal cumulative distribution function of $F_n$ with respect to the components with indices ${\mathcal J}$.

Hence, the resulting tests are $$\varphi_\ell := \mathbbm{1}\left\{\forall \mathcal{J}\subset \{1,\ldots,L\} \text{ with } \ell \in \mathcal{J}: \:  \exists\, {j \in\mathcal{J}}: \: {W_{j,n}} > q_{j,n}^{\mathcal J} \right\}$$ for $\mathcal{H}_{0,\ell}, \ell\in\{1,\ldots,L\},$ and $\varphi := \max_{\ell \in \{1,\ldots,L\}} \varphi_\ell$ for $\mathcal H_0$.
Under the same assumptions as in Theorem~\ref{BSQuant}, these tests fulfill \eqref{eq:(mid)},
\begin{align*}
    &\limsup_{n\to\infty} \E^*\left(\max_{\ell\in\T} \varphi_\ell\right) \leq \alpha &\text{ under $\bigcap_{\ell\in\T} \mathcal{H}_{0,\ell}$ for all index sets $\T\subset \{1,\dots,L\}$, and}\\
    &\liminf_{n\to\infty}\E^*\left(\varphi_\ell\right) \geq \mathrm{FWER}^{-1}(\alpha) &\text{ under $\mathcal{H}_{0,\ell}$ for all $\ell\in\{1,\dots,L\}$}
\end{align*}as shown in the following:\medskip

Let $\T\subset \{1,\dots,L\}$ be an index set of true hypotheses.
Under $\bigcap_{\ell\in\T} \mathcal{H}_{0,\ell}$, we have for all $\ell \in \T$
\begin{align*}
    \varphi_\ell &= \mathbbm{1}\left\{\forall \mathcal{J}\subset \{1,\ldots,L\} \text{ with } \ell \in \mathcal{J}: \:  \exists \, {j \in\mathcal{J}}: \: {W_{j,n}} > {q_{j,n}^{\mathcal J}}  \right\}\\
    &\leq \mathbbm{1}\left\{\exists \, {j \in\mathcal{T}}: \: {W_{j,n}} > {q_{j,n}^{\mathcal T}} \right\}
    \\&= \max_{j \in\mathcal{T}} \mathbbm{1}\left\{{W_{j,n}} > {q_{j,n}^{\mathcal T}}  \right\}
\end{align*} 
and
\begin{align*}
    \varphi_\ell &= \mathbbm{1}\left\{\forall \mathcal{J}\subset \{1,\ldots,L\} \text{ with } \ell \in \mathcal{J}: \:  \exists \, {j \in\mathcal{J}}: \: {W_{j,n}} > {F_{j,n}^{-1}\left( 1-\beta_n^{\mathcal J} \right)} \right\}\\
    &\geq \mathbbm{1}\left\{\forall \mathcal{J}\subset \{1,\ldots,L\} \text{ with } \ell \in \mathcal{J}: \:   \: {W_{\ell,n}} > {F_{\ell,n}^{-1}\left( 1-\beta_n^{\mathcal J} \right)} \right\}
    \\&\geq \mathbbm{1}\left\{{W_{\ell,n}} > {F_{\ell,n}^{-1}\left( 1-\mathrm{FWER}_{n}^{-1}(\alpha)  \right)}  \right\}&\text{($\beta_n^{\mathcal{J}} \geq 
    \mathrm{FWER}_{n}^{-1}(\alpha) $ by definition)}
    \\&= \mathbbm{1}\left\{{W_{\ell,n}} > q_{\ell,n}  \right\}
    .
\end{align*} 
Hence, it follows
\begin{align*}
    \E^*\left(\max_{\ell\in\T}\mathbbm{1}\left\{{W_{\ell,n}} > q_{\ell,n}  \right\}\right)
    &\leq \E^*\left(\max_{\ell\in\mathcal{T}} \varphi_\ell\right)
    \leq \E^*\left(\max_{j \in\mathcal{T}} \mathbbm{1}\left\{{W_{j,n}} > {q_{j,n}^{\mathcal T}}  \right\}\right).
\end{align*}
Now, we can apply Theorem~\ref{BSQuant} to obtain \eqref{eq:(start)} for the non-extended tests $$\mathbbm{1}\left\{{W_{\ell,n}}>q_{\ell,n} \right\},\ell\in\{1,\ldots,L\},\quad \text{ and }\quad\mathbbm{1}\left\{{W_{\ell,n}}>q_{\ell,n}^{\mathcal T}  \right\}, \ell\in\T,$$
to prove that the lower and upper bound converge to values less or equal than $\alpha$.
If $\T = \{1,\ldots,L\}$, the lower and upper bound both converge to $\alpha$ by \eqref{eq:(mid)}.
If $\T = \{\ell\}$ for some $\ell\in\{1,\ldots,L\}$, the lower bound converges to $\mathrm{FWER}^{-1}(\alpha)$ by \eqref{eq:(end)} and the upper bound converges to $\alpha$ by \eqref{eq:(mid)}.
\hfill\qed

\section{Proofs}
This section contains the proofs of Theorem~\ref{BSQuant} and Theorem~\ref{pValues}.

\medskip
\noindent\textbf{Proof of Theorem~\ref{BSQuant}}
For proving Theorem~\ref{BSQuant}, we firstly need two lemmas. The first one is a multivariate version of Polya's theorem.

\begin{lem}\label{versionofS4}
    Let $L\in\mathbb N$, $A := \bigtimes_{\ell=1}^L [a_\ell,b_\ell] \subset \bar{\mathbb R}^L := (\R \cup \{-\infty,\infty\})^L$,
    $\mathcal{F}\in \mathrm{CDF}(\R^L)$ be a cumulative distribution function that is continuous on $A$ and $({F}_n(\mathbf t))_{\mathbf t\in\R^L},n\in\mathbb N,$ denote a sequence of maps on a probability space $(\Omega,\mathcal A, \P)$ taking values in the space of all cumulative distribution functions $\mathrm{CDF}(\R^L)$ satisfying \begin{align}
        \label{eq:condition3}
        F_{n}(\mathbf{t}) \xrightarrow{\P} \mathcal F (\mathbf{t}) \text{ as } n\to\infty \text{ for all } \mathbf{t}\in A.
    \end{align} Here, we extend a cumulative distribution function to $\bar \R^L$ by taking coordinate-wise limits. Then, we have
$$ \sup\limits_{\mathbf{t}\in A} |{F}_n(\mathbf{t}) - \mathcal{F} (\mathbf{t})| \xrightarrow{\P} 0 \quad\text{as } n\to\infty.$$
\end{lem}
\begin{proof}[Proof of Lemma~\ref{versionofS4}]
We aim to apply Proposition~2.1 in \cite{multivariatePolya}.
    Therefore, let $f_{\mathbf{t}}:\R^L \to \R, f_{\mathbf{t}}(x_1,\ldots,x_L) = \mathbbm{1}\{ x_1 \leq t_1, \ldots, x_L \leq t_L \}$ for all $\mathbf{t} = (t_1,\ldots,t_L)^{\top} \in A$,
    $\mathbf{F} := \left\{ f_{\mathbf{t}}  \mid \mathbf{t}  \in A \right\}$ and $\varepsilon > 0$ be arbitrary. Furthermore, let $\mathcal F_1,\ldots,\mathcal F_L:\bar\R \to [0,1]$ denote the (extended) marginal cumulative distribution functions of $\mathcal{F}$,
    $m\in\mathbb{N}$ with $\varepsilon/L \geq 1/m$ and define $$a_\ell =: t_{\ell, 0}  < t_{\ell, 1} < \ldots  < t_{\ell,m} := b_\ell$$ such that
    $\mathcal{F}_{\ell} (t_{\ell,i}) - \mathcal{F}_{\ell} (t_{\ell,i-1}) \leq \varepsilon/L$ for all $i\in\{1,\ldots,m\}, \ell\in\{1,\ldots,L\}$.
    Set $\mathbf{t}_{i_1,\ldots,i_L} := (t_{1,i_1},\ldots,t_{L,i_L})^{\prime}$ for all $i_1,\ldots,i_L\in\{0,\ldots,m\}$.
    Then, it holds that
    \begin{align*}
        \int_{\R^L}\left( f_{\mathbf{t}_{i_1,\ldots,i_L}} - f_{\mathbf{t}_{i_1-1,\ldots,i_L-1}}\right)\;\mathrm{d}\mathcal{F} &= 
        \mathcal{F}(\mathbf{t}_{i_1,\ldots,i_L}) - \mathcal{F}(\mathbf{t}_{i_1-1,\ldots,i_L-1})
        \leq 
        \sum\limits_{\ell =1}^L \left( \mathcal{F}_{\ell} (t_{\ell,i_\ell}) - \mathcal{F}_{\ell} (t_{\ell,i_\ell-1})\right)
        \leq \varepsilon
    \end{align*}
    for all $i_1,\ldots,i_L\in\{1,\ldots,m\}$. Thus, the bracketing number is bounded by $(m+1)^L < \infty$. 
As in the proof of Proposition~2.1 in \cite{multivariatePolya}, 
it holds
\begin{align*}
    \sup_{\mathbf{t}\in A} |{F}_n(\mathbf{t}) - \mathcal{F}(\mathbf{t})|
    \leq 3\max\limits_{i_1,\ldots,i_L \in \{0,\ldots,m\}}\{|{F}_n(\mathbf{t}_{i_1,\ldots,i_L}) - \mathcal{F}(\mathbf{t}_{i_1,\ldots,i_L})|\} + 2\varepsilon.
\end{align*}
 Hence, 
 we get \begin{align*}
     \P^*\left( \sup_{\mathbf{t}\in A} |{F}_n(\mathbf{t}) - \mathcal{F}(\mathbf{t})| > 5\varepsilon \right)
     \leq \P^*\left( \max\limits_{i_1,\ldots,i_L \in \{0,\ldots,m\}}\{|{F}_n(\mathbf{t}_{i_1,\ldots,i_L}) - \mathcal{F}(\mathbf{t}_{i_1,\ldots,i_L})|\} > \varepsilon  \right),
 \end{align*}
 where $\P^*$ denotes the outer probability.
 The latter tends to 0 by \eqref{eq:condition3} since the maximum is taken over a finite number of values. 
\end{proof}

The following lemma ensures that the quantiles of a converging sequence of cumulative distribution functions converge uniformly.
\begin{lem}\label{Contlem}
Let $\mathcal{F}\in\mathrm{CDF}(\R)$ be a cumulative distribution function that is continuous and strictly increasing on $[a,b] \subset \R$ and $\left({F}_n(t)\right)_{t\in\R}, {n\in\mathbb N},$ denote a sequence of maps on a probability space $(\Omega,\mathcal A, \P)$ taking values in the space of all cumulative distribution functions $\mathrm{CDF}(\R)$ satisfying 
\begin{align}\label{eq:ptwGn}
      {F}_n(t) \xrightarrow{P} \mathcal{F}(t) \quad \text{ for all } t\in [a,b]
\end{align}
 as $n\to\infty$.
Furthermore, let $\mathcal{F}(a) < p \leq q < \mathcal{F}(b).$ Then, we have
$$ \sup\limits_{r \in [p,q]} | {F}_n^{-1}(r) - \mathcal{F}^{-1}(r) | \xrightarrow{\P} 0 $$ as $n\to\infty$.
\end{lem}
\begin{proof}[Proof of Lemma~\ref{Contlem}]
Let $\varepsilon > 0$ be arbitrary. We aim to show
\begin{align*}
    P^*\left( \sup\limits_{r \in [p,q]} | {F}_n^{-1}(r) - \mathcal{F}^{-1}(r) | > \varepsilon \right) \to 0 \quad\text{as } n\to\infty.
\end{align*}
Since $\mathcal F$ is strictly increasing on $[a,b]$, the inverse function $\mathcal F^{-1}: [0,1] \to \bar\R$ is continuous on the compact interval $\left[ \frac{\mathcal F(a) +p}{2}, \frac{q + \mathcal F(b)}{2} \right]$ and, thus, uniformly continuous on that interval.
Hence, there exists a $\delta > 0$ such that
\begin{align}
    \label{eq:uni1}
 \forall r_1,r_2 \in \left[ \frac{\mathcal F(a) +p}{2}, \frac{q + \mathcal F(b)}{2} \right]  :\quad  |r_1-r_2| \leq\delta \quad\Rightarrow \quad | \mathcal F^{-1}(r_1) - \mathcal F^{-1}(r_2) | \leq \varepsilon.
\end{align}
Let $a':= \mathcal{F}^{-1}\left( \frac{\mathcal F(a) +p}{2}\right) > a,b' := \mathcal{F}^{-1}\left( \frac{q + \mathcal F(b)}{2} \right) < b.$
Moreover, $\mathcal F$ is continuous on $[a,b]$. Hence, it is uniformly continuous on $[a,b]$. Thus, there exists a $\delta' \in \left( 0  , b-a \right]$ such that $$ | \mathcal F(t_1) - \mathcal F(t_2) | \leq \delta/2 $$ holds for all $t_1,t_2 \in \left[ a,b \right]$ with $|t_1-t_2| \leq\delta'$.
Additionally, let $(\delta_n)_{n\in\mathbb N}$ be a positive sequence with $\delta_n \to 0$ as $n\to\infty$. Then, let $N\in\mathbb N$  such that $\delta_n < \min\{a'-a,b-b',\delta'\}$ for all $n \geq N$.

By using $\mathcal F^{-1}(\mathcal F(t)) = t$ for $t\in [a,b)$ and \eqref{eq:uni1}, we observe that
\begin{align*}
    &P^*\left( \sup\limits_{r \in [p,q]} | {F}_n^{-1}(r) - \mathcal{F}^{-1}(r) | > \varepsilon \right) 
    \\& \leq P^*\left( \sup\limits_{r \in [p,q]} | \mathcal{F}^{-1}\left(\mathcal F \left({F}_n^{-1}(r)\right)\right) - \mathcal{F}^{-1}(r) | > \varepsilon \:\lor\: \exists r\in[p,q]:\:{F}_n^{-1}(r) \not\in [a,b) \right) 
     \\& \leq P^*\left( \exists {r \in [p,q]}:\: |\mathcal F \left({F}_n^{-1}(r)\right) - r | > \delta \:\lor\: \exists r\in[p,q]:\:{F}_n^{-1}(r) \not\in [a',b'] \right) 
\end{align*} for all $n\in\mathbb N$.
Now, we can use that $F_n(F_n^{-1}(r)) \geq r$ and $F_n(F_n^{-1}(r) - \delta_n) \leq r$ for all $n\in\mathbb N, r \in [p,q]$ due to the definition of the upper quantile.
Hence, it follows
\begin{align*}
    &P^*\left( \sup\limits_{r \in [p,q]} | {F}_n^{-1}(r) - \mathcal{F}^{-1}(r) | > \varepsilon \right) 
     \\& \leq P^*\Big( \exists {r \in [p,q]}:\:  \mathcal F \left({F}_n^{-1}(r)\right) -F_n(F_n^{-1}(r) - \delta_n)  > \delta \:\lor\: F_n(F_n^{-1}(r)) - \mathcal F \left({F}_n^{-1}(r)\right)  > \delta \\ &\qquad\qquad \lor\: \exists r\in[p,q]:\:{F}_n^{-1}(r) \not\in [a',b'] \Big) 
     \\& \leq P^*\Big( \exists {r \in [p,q]}:\:  \mathcal F \left({F}_n^{-1}(r)\right) - \mathcal F(F_n^{-1}(r)- \delta_n) + \mathcal F(F_n^{-1}(r) - \delta_n)-F_n(F_n^{-1}(r) - \delta_n)  > \delta\\ &\qquad\qquad \:\lor\: \sup_{t\in[a,b]} |F_n(t) - \mathcal F \left(t\right)|  > \delta  \:\lor\: \exists r\in[p,q]:\:{F}_n^{-1}(r) \not\in [a',b'] \Big) \end{align*} for all $n\in\mathbb N$.
Note that for $n\geq N$, we have that $\delta_n < \min\{a'-a,b-b',\delta'\}$ and, thus, $ \mathcal F \left(t\right) - \mathcal F(t- \delta_n) \leq\delta/2 $ for all $t\in [a',b]$.
Hence, we obtain
     \begin{align*}
    &P^*\left( \sup\limits_{r \in [p,q]} | {F}_n^{-1}(r) - \mathcal{F}^{-1}(r) | > \varepsilon \right)
     \\& \leq P^*\Big( \exists {r \in [p,q]}:\:  \delta/2 + \mathcal F(F_n^{-1}(r) - \delta_n)-F_n(F_n^{-1}(r) - \delta_n)  > \delta\\ &\qquad\qquad \:\lor\: \sup_{t\in[a,b]} |F_n(t) - \mathcal F \left(t\right)|  > \delta  \:\lor\: \exists r\in[p,q]:\:{F}_n^{-1}(r) \not\in [a',b'] \Big) 
      \\& \leq P^*\left( \sup_{t\in [a,b]}| F_n(t) - \mathcal F(t)|  > \delta/2 \:\lor\: \exists r\in[p,q]:\:{F}_n^{-1}(r) \not\in [a',b'] \right) 
       \\& \leq P^*\left( \sup_{t\in [a,b]}| F_n(t) - \mathcal F(t)|  > \delta/2 \right) + P^*\left(  \exists r\in[p,q]:\:{F}_n^{-1}(r) \not\in [a',b'] \right) 
\end{align*} for all $n\geq N$.
The first summand converges to $0$ by Lemma~\ref{versionofS4}.
For the second summand $$ P^*\left(  \exists r\in[p,q]:\:{F}_n^{-1}(r) \not\in [a',b'] \right) 
  = P^*\left(  {F}_n^{-1}(p) < a' \:\lor\: {F}_n^{-1}(q)> b'\right) ,$$ we use that
  ${F}_n^{-1}(q)> b'$ is equivalent to $q\geq F_n( b')$ and that ${F}_n^{-1}(p) < a'$ implies $p < F_n(a')$. Then, we have
\begin{align*}
  P^*\left(  \exists r\in[p,q]:\:{F}_n^{-1}(r) \not\in [a',b'] \right)
  &\leq P^*\left(  p < F_n(a') \:\lor\: q\geq F_n( b')\right)
  \\&\leq P^*\left(  p < F_n(a') \right) + P^*\left( q\geq F_n( b')\right)
   \\&\leq P^*\left(  p - \mathcal F(a') < F_n(a')- \mathcal F(a') \right) + P^*\left(  \mathcal F(b') -q\leq  \mathcal F(b') - F_n( b')\right)
   \\& \to 0
\end{align*} as $n\to\infty$ by \eqref{eq:ptwGn} since $p > \mathcal  F(a')$ and $\mathcal F(b') > q$.
This completes the proof.
\end{proof}

    Now, we can prove Theorem~\ref{BSQuant}.
    We start by showing that $\beta_n$ converges in outer probability.  Therefore, it remains to show 
    \begin{align}\label{eq:FWERQuantil}
         \left|\mathrm{FWER}_{n}^{-1}(\alpha) - \mathrm{FWER}^{-1}(\alpha)\right|
        \xrightarrow{P} 0\quad \text{ as } n\to\infty
    \end{align} by applying Lemma~\ref{Contlem}. Note that $\mathrm{FWER}_{n}$ and $\mathrm{FWER}$ can be seen as left-continuous cumulative distribution functions and $\mathrm{FWER}$ is continuous and strictly increasing on $[a,b]$ by assumption.
    By Lemma~\ref{versionofS4}, we have 
    $$\sup\limits_{\mathbf t\in\R^L} |F_n(\mathbf t) - \mathcal{F}(\mathbf t)| \xrightarrow{P} 0 \quad \text{ as } n\to\infty.$$
    Moreover, Lemma~\ref{versionofS4} implies the uniform convergence of the marginal cumulative distribution functions, that is
    \begin{align}\label{eq:Flconv}
        \sup\limits_{t\in\R} |F_{\ell,n}(t) - \mathcal F_{\ell}(t)|
        \xrightarrow{P} 0
    \end{align} as $n\to\infty$ for all $\ell\in\{1,\ldots,L\}$. 
Furthermore, we have
\begin{align*}
   \sup_{\zeta\in [0,1]} \left| \mathrm{FWER}_{n}(\zeta) - \mathrm{FWER}(\zeta)\right| &=  \sup_{\zeta\in [0,1]}
    \left|F_n\left( F_{1,n}^{-1}(1-\zeta),\dots,F_{L,n}^{-1}(1-\zeta)  \right) - \mathcal F\left( \mathcal F_{1}^{-1}(1-\zeta),\dots,\mathcal F_{L}^{-1}(1-\zeta)  \right)\right|
    \\&\leq \sup\limits_{\mathbf t\in\R^L} |F_n(\mathbf t) - \mathcal{F}(\mathbf t)|
    \\&\quad+  \sup_{\zeta\in [0,1]}\left|\mathcal F\left( F_{1,n}^{-1}(1-\zeta),\dots,F_{L,n}^{-1}(1-\zeta)  \right) -\mathcal F\left( \mathcal F_{1}^{-1}(1-\zeta),\dots,\mathcal F_{L}^{-1}(1-\zeta)  \right)\right|.
\end{align*}
Here, the first summand converges to zero in outer probability. 
For the second summand, note that 
\begin{align}
    &\mathcal F_\ell\left(\mathcal F_{\ell}^{-1}(1-\zeta)  \right) = 1-\zeta \text{ and}\label{eq:i}\\
    &F_-(w) \leq 1-\zeta\Leftrightarrow w\leq F^{-1}(1-\zeta) \label{eq:ii}
\end{align}
 hold for all $\zeta\in [0,1], w\in\R$ and all cumulative distribution functions $F\in\mathrm{CDF}(\R)$ with left-continuous version $F_-$.
Let $\varepsilon > 0$ and $\ell\in\{1,\ldots,L\}$ be arbitrary.
Then, it holds
\begin{align*}
    \mathcal F_\ell\left(F^{-1}(1-\zeta)  \right) &= \P\left( W_\ell \leq F^{-1}(1-\zeta) \right)
     = \P\left( F_-(W_\ell) \leq 1-\zeta \right)
    \begin{cases}
        \leq \P\left( \mathcal F_{\ell}(W_\ell) \leq 1-\zeta+\varepsilon \right) = 1-\zeta+\varepsilon \\
        \geq \P\left( \mathcal F_{\ell}(W_\ell) \leq 1-\zeta -\varepsilon \right) = 1-\zeta-\varepsilon
    \end{cases}
\end{align*}
by \eqref{eq:i} and \eqref{eq:ii} for all cumulative distribution functions $F:\R\to\R$ with $\sup_{t\in\R}|F(t) - \mathcal{F}_\ell(t)| \leq \varepsilon$ and all $\zeta \in [0,1]$.
Thus, it follows
$$ \sup_{\zeta\in [0,1]} |\mathcal F_\ell\left(F^{-1}(1-\zeta)  \right) - \mathcal F_\ell\left(\mathcal F_{\ell }^{-1}(1-\zeta)  \right) | \leq \varepsilon $$ by \eqref{eq:i} for all cumulative distribution functions $F:\R\to\R$ with $\sup_{t\in\R}|F(t) - \mathcal{F}_\ell(t)|\leq\varepsilon$.
Hence, \eqref{eq:Flconv} implies
\begin{align*}
    \P^*\left( \sup_{\zeta\in [0,1]} |\mathcal F_\ell\left(F_{\ell,n}^{-1}(1-\zeta)  \right) - \mathcal F_\ell\left(\mathcal F_{\ell }^{-1}(1-\zeta)  \right) | > \varepsilon \right) \leq \P^*\left( \sup_{t\in\R}|F_{\ell,n}(t) - \mathcal{F}_\ell(t)| > \varepsilon \right) \to 0
\end{align*} as $n\to\infty$.
Thus, by $$|\mathcal F(x_1,\ldots,x_L)  - \mathcal F(y_1,\ldots,y_L) | \leq \sum_{\ell=1}^L |\mathcal F_\ell(x_\ell)  - \mathcal F_\ell(y_\ell) |$$ for all $x_1,\ldots,x_L, y_1,\ldots,y_L\in\R$, the second summand converges to zero in outer probability as well.
    Hence, it follows
    \begin{align*}
        \sup_{\zeta\in [0,1]}\left| \mathrm{FWER}_{n}(\zeta) - \mathrm{FWER}(\zeta)\right| \xrightarrow{P} 0 \quad \text{as $n\to\infty$}.  
    \end{align*} 
    Thus, for the right-continuous version $\mathrm{FWER}_{n+}$ of $\mathrm{FWER}_{n}$, we obtain
    \begin{align*}
         \sup_{\zeta\in [0,1)}\left| \mathrm{FWER}_{n+}(\zeta) - \mathrm{FWER}(\zeta)\right| \xrightarrow{\P} 0
    \end{align*}
     as $n\to\infty$.
    Applying Lemma~\ref{Contlem} yields \eqref{eq:FWERQuantil} and, thus, $\beta_n \xrightarrow{P} \mathrm{FWER}^{-1}(\alpha)$ as $n\to\infty$.

In order to proof the statements \eqref{eq:(i')}--\eqref{eq:(ii')}, let $\mathcal{T}\subset \{1,\ldots,L\}$ be arbitrary and $P \in \bigcap_{\ell\in\T} \mathcal P_{\ell}$, i.e., $\mathcal H_{0,\ell}, \ell\in\mathcal T,$ be true.
Note that \eqref{eq:ii} implies
\begin{align}\label{eq:eqeq}
    W_{\ell,n} > q_{\ell,n}   \quad\Leftrightarrow\quad F_{\ell,n-}(W_{\ell,n}) > 1 - \beta_n  \qquad \text{for all $\ell\in\{1,\ldots,L\}.$}
\end{align}
By $\beta_n \xrightarrow{P} \mathrm{FWER}^{-1}(\alpha)$ as $n\to\infty$, it follows $$\left(\left(W_{\ell,n} \right)_{\ell\in\T},\beta_n \right)\xrightarrow{d} \left( \left(W_{\ell} \right)_{\ell\in\T}, \mathrm{FWER}^{-1}(\alpha)\right)$$ as $n\to\infty$ by \eqref{eq:DistrConv} and Slutsky's theorem. 
Moreover, the continuous mapping theorem implies
$$\left(\left(\mathcal F_\ell(W_{\ell,n})\right)_{\ell\in\T}, \beta_n\right) \xrightarrow{d} \left(\left(\mathcal F_\ell(W_{\ell})\right)_{\ell\in\T}, \mathrm{FWER}^{-1}(\alpha)\right)$$ as $n\to\infty$ due to the continuity of $\mathcal F_1,\ldots,\mathcal F_L$.
By \eqref{eq:Flconv}, we have
$$| F_{\ell,n-}(W_{\ell,n}) - \mathcal F_\ell(W_{\ell,n})| \leq \sup_{t\in\R} |F_{\ell,n}(t) - \mathcal F_\ell(t)| \xrightarrow{\P} 0$$ for all $\ell\in\T$ and, thus,
Slutsky's theorem implies
$$\left(\left( F_{\ell,n-}(W_{\ell,n})\right)_{\ell\in\T}, \beta_n\right) \xrightarrow{d} \left(\left(\mathcal F_\ell(W_{\ell})\right)_{\ell\in\T}, \mathrm{FWER}^{-1}(\alpha)\right)$$ as $n\to\infty$.

Now, by approximating the function $$f:\R^{|\T|}\times \R \to \{0,1\}, \quad f ((t_\ell)_{\ell\in\T},b)=\max_{\ell\in\T}\mathbbm{1}\{t_\ell > 1-b\} $$ by sequences of Lipschitz functions $(g_m)_{m\in\mathbb N}, (h_m)_{m\in\mathbb N}$ with $1\geq g_m \geq f \geq h_m \geq 0$ and $$\E\left[g_m\left((\mathcal F_\ell({W}_\ell))_{\ell\in\T},\mathrm{FWER}^{-1}(\alpha)\right) - h_m\left(\mathcal F_\ell({W}_\ell))_{\ell\in\T},\mathrm{FWER}^{-1}(\alpha)\right)\right] \leq m^{-1}$$ for all $m\in\mathbb{N}$, one can follow similarly as in the proof of Lemma~\ref{MonteCarloRemark} that \begin{align*}
    \P^*\left(\exists \ell \in \T:\: F_{\ell,n-}(W_{\ell,n}) > 1-\beta_n\right) &= \E^* \left[f\left((F_{\ell,n-}(W_{\ell,n}))_{\ell\in\T}, \beta_n\right)\right] \\&\to \E \left[ f\left((\mathcal F_\ell(W_{\ell}))_{\ell\in\T}, \mathrm{FWER}^{-1}(\alpha)\right)\right] \\&= \P(\exists \ell \in \T:\: \mathcal F_\ell(W_{\ell}) > 1-\mathrm{FWER}^{-1}(\alpha)) \\&\leq 1 - \P(\forall \ell \in \{1,\dots,L\}:\: W_{\ell} \leq \mathcal F_{\ell}^{-1}(1-\mathrm{FWER}^{-1}(\alpha))) \\&= 
    \mathrm{FWER}(\mathrm{FWER}^{-1}(\alpha)) = \alpha 
 \end{align*} holds as $n\to\infty$  under $\bigcap_{\ell\in\T}\mathcal{H}_{0,\ell}$, yielding \eqref{eq:(i')} by \eqref{eq:eqeq}.
The inequality is an equality if $\T = \{1,\ldots,L\}$, yielding \eqref{eq:(i'')} by \eqref{eq:eqeq}.
Moreover, we observe that 
\begin{align*}
    \P^*\left( F_{\ell,n-}(W_{\ell,n}) > 1-\beta_n\right) &\to 
    \P( W_{\ell} > \mathcal{F}_{\ell}^{-1}( 1-\mathrm{FWER}^{-1}(\alpha)) ) 
      \\&= 1 - \mathcal{F}_\ell(\mathcal{F}_{\ell}^{-1}( 1-\mathrm{FWER}^{-1}(\alpha))) = \mathrm{FWER}^{-1}(\alpha)
 \end{align*}
follows as $n\to\infty$ for $\T = \{\ell\}$ for some $\ell \in\{1,\ldots,L\}$, yielding \eqref{eq:(ii')} by \eqref{eq:eqeq}.
\hfill\qed

\begin{rem}\label{RemarkBSQuant}
    It results from the proof 
    that  $\beta_n \xrightarrow{\P} \mathrm{FWER}^{-1}(\alpha)$ as $n\to\infty$.
\\
    Furthermore, by applying Lemma~\ref{Contlem}, one can show $q_{\ell, n} \xrightarrow{P} \mathcal F_\ell^{-1}(1-\mathrm{FWER}^{-1}(\alpha))$ as $n\to\infty$ for all $\ell\in\{1,\ldots,L\}$ if $\mathcal F_\ell$  is strictly increasing on $[\widetilde a, \widetilde b]$  with $\mathcal F_\ell (\widetilde a) < 1-\alpha \leq 1-\alpha/L < \mathcal F_\ell (\widetilde b)$  for all $\ell\in\{1,\ldots,L\}$.    
\end{rem}

\begin{rem}\label{LowerQuant}
    Note that the critical values $q_{\ell,n}$ can also be chosen as any other $(1-\beta_n)$-quantile of $F_{\ell,n}$ instead of the upper quantiles for all $\ell\in\{1,\ldots,L\}$.
    To see this, let $q'_{\ell,n} \leq q_{\ell,n}$ be a $(1-\beta_n)$-quantile of $F_{\ell,n}$. 
    Then, for all $\mathcal T \subset \{1,\ldots,L\}$,
    \begin{align*}
        \P^*\left( \exists \ell \in\mathcal T:
        W_{\ell,n} > q'_{\ell,n}\right)
        \geq \P^*\left( \exists \ell \in\mathcal T:
        W_{\ell,n} > q_{\ell,n}\right),
    \end{align*} which leads lower bounds.
    On the other hand,
    \begin{align*}
        \P^*\left( \exists \ell \in\mathcal T:
        W_{\ell,n} > q'_{\ell,n}\right)
        \leq \P^*\left( \exists \ell \in\mathcal T:
        F_{\ell,n-}(W_{\ell,n}) \geq 1-\beta_n\right).
    \end{align*}
    Similarly as in the preceding proof, the latter converges to 
    \begin{align*}
         \P^*\left( \exists \ell \in\mathcal T:
       \mathcal F_{\ell-}(W_{\ell}) \geq 1-\mathrm{FWER}^{-1}(\alpha)\right) = \P^*\left( \exists \ell \in\mathcal T:
        \mathcal F_{\ell}(W_{\ell}) > 1-\mathrm{FWER}^{-1}(\alpha)\right)
    \end{align*} under $\bigcap_{\ell\in\mathcal T}\mathcal H_{0,\ell}$ due to the continuity of $\mathcal F_\ell$.
    Hence, $\P^*( \exists \ell \in\mathcal T:
        W_{\ell,n} > q'_{\ell,n})$ converges to the same value as $\P^*\left( \exists \ell \in\mathcal T:
        W_{\ell,n} > q_{\ell,n}\right)$  under $\bigcap_{\ell\in\mathcal T}\mathcal H_{0,\ell}$.

        However, in this work, we focus on upper quantiles, as those provide more intuitive critical values in the case of non-continuous cumulative distribution functions.
        To be concrete, choosing upper quantiles $q_{1,n},\ldots,q_{L,n}$ ensures that $1 -F_n(q_{1,n},\ldots,q_{L,n}) \leq \alpha$ for finite $n\in\mathbb N$.
\end{rem}

\medskip
\noindent\textbf{Proof of Theorem~\ref{pValues}}
   For (1), let $\ell\in\{1,\ldots,L\}$ be fixed.
       Firstly, we aim to show $$p_{\ell} \leq \alpha \quad \Rightarrow \quad W_{\ell,n} > q_{\ell,n}.$$ Therefore, assume that $p_{\ell} \leq \alpha$ holds. Since $p_{\ell} \leq \alpha$ implies that $\beta_{\ell,n}$ satisfies
       $ \mathrm{FWER}_{n}(\beta_{\ell,n}) \leq \alpha , $
        it follows $$1 - F_{\ell,n -}(W_{\ell,n})+ B_n^{-1} = \beta_{\ell,n}  \leq \beta_n$$ by the definition of $\beta_n$ and, thus, $1 - F_{\ell,n -}(W_{\ell,n}) < \beta_n$. 
        Hence, we have $W_{\ell,n} > F_{\ell,n }^{-1}(1-\beta_n) = q_{\ell,n}$
        by \eqref{eq:eqeq}.\\
        Secondly,  we aim to prove $$W_{\ell,n} > q_{\ell,n} \quad \Rightarrow \quad p_{\ell} \leq \alpha.$$
        The inequality $W_{\ell,n} > q_{\ell,n} = F_{\ell,n}^{-1}\left( 1-\beta_n \right)$ implies 
        \begin{align*}
            \beta_{\ell,n} = 1 - F_{\ell,n -}(W_{\ell,n}) + B_n^{-1} \leq 1 - F_{\ell,n}(F_{\ell,n}^{-1}\left( 1-\beta_n \right)) + B_n^{-1}< \beta_n + B_n^{-1} .
        \end{align*} Due to $\beta_{\ell,n}, \beta_n \in\{0,\frac{1}{B_n},\frac{2}{B_n},\ldots,1\}$, it follows $\beta_{\ell,n} \leq \beta_n $.
        Thus, the definition of $\beta_n$ yields that $\beta_{\ell,n}$ fulfills
        $$ p_{\ell} = \mathrm{FWER}_{n}(\beta_{\ell,n}) \leq \alpha . $$

        For (2), we note that $p = \min\{p_1,\ldots,p_L\} \leq \alpha$ if and only if there exists an $\ell\in\{1,\ldots,L\}$ such that $p_{\ell} \leq \alpha$. Due to (1), this holds whenever there exists an $\ell\in\{1,\ldots,L\}$ such that $W_{\ell,n} > q_{\ell,n}$.
\hfill\qed

\end{document}